\documentclass[12pt,journal,onecolumn]{IEEEtran}

\usepackage{amsmath, amsfonts, amssymb, amsopn}
\usepackage{pifont, mathrsfs, pstricks}
\usepackage{citesort}
\usepackage{research5}
\usepackage{graphicx,epsfig}
\usepackage{psfrag}

\newtheorem{lemma}{Lemma}[section]
\newtheorem{corollary}{Corollary}[section]
\newtheorem{proposition}{Proposition}[section]
\newtheorem{definition}{Definition}[section]

\begin{document}

\title{Low SNR Capacity of Noncoherent Fading Channels}
\author{Vignesh Sethuraman, Ligong Wang,~\IEEEmembership{student member,~IEEE}, Bruce~Hajek,~\IEEEmembership{Fellow,~IEEE} 
and~Amos~Lapidoth,~\IEEEmembership{Fellow,~IEEE}
\thanks{V. Sethuraman is with Qualcomm, Santa Clara, California.
L. Wang and A. Lapidoth are with the Signal and Information Processing Laboratory, ETH Zurich.
B. Hajek is with the Department of Electrical and Computer Engineering and the Coordinated
Science Laboratory at the University of Illinois. 
This work was conducted while V. Sethuraman was at the University of Illinois.}
\thanks{This work was supported in part by the National Science Foundation Grant NSF ITR 00-85929, and the Motorola Center for Communication Graduate Fellowship.}
\thanks{Portions of this work appeared in \cite{VigneshBH06,LapidothWang06,VWHL07}.}
}
\maketitle

\begin{abstract}
Discrete-time Rayleigh fading single-input single-output (SISO) and multiple-input multiple-output (MIMO) channels are considered, with no channel state information at the transmitter or the receiver. The fading is assumed to be stationary and correlated in time, but independent from antenna to antenna. Peak-power and average-power constraints are imposed on the transmit antennas. For MIMO channels, these constraints are either imposed on the sum over antennas, or on each individual antenna. For SISO channels and MIMO channels with sum power constraints, the asymptotic capacity as the peak signal-to-noise ratio goes to zero is identified; for MIMO channels with individual power constraints, this asymptotic capacity is obtained for a class of channels called transmit separable channels. The results for MIMO channels with individual power constraints are carried over to SISO channels with delay spread (i.e. frequency selective fading).   
\end{abstract}



\section{Introduction}
In this paper we present results on the capacity of discrete-time Rayleigh fading single-input single-output (SISO) and multiple-input multiple-output (MIMO) channels. We assume a noncoherent  model where no channel state information is available at the transmitter or the receiver. The fadings are assumed to be stationary processes correlated in time but, for MIMO channels, independent for distinct input/output antenna pairs. A hard peak-power constraint, in addition to an average-power constraint, is imposed on the input signals. For MIMO channels we consider two types of constraints: under one the peak and average power constraints are imposed on each of the signals transmitted by the different antennas separately, and  under the other the constraints are on the sum of the powers in the different signals. We focus on channel capacity at low signal-to-noise ratio (SNR), but we also derive upper bounds that are valid for any SNR.

We also consider SISO channels with delay spread (i.e., frequency selective) fading where the fading is modeled by a finite number of taps. The fading processes corresponding to the different taps are assumed to be independent across taps, and allowed, within each tap, to be correlated in time. 

The capacity of fading channels at low SNR was studied in \cite{MedardGallager02,SubramanianHajek02,HajekSubramanian02,PrelovVerdu04,Laneman05,Collins05,RaoHassibi04,SrinivasanVaranasi06,WuSrikant07,DurisiBoelcskeiShamai06,VigneshHajek05,VigneshBH05,Jacobs63,Kennedy69,TelatarTse00}. 
The main motivation for our present work has been to understand the capacity of communication
over wideband channels.  Work of Kennedy \cite{Kennedy69}, Jacobs \cite{Jacobs63},  
Telatar and Tse \cite{TelatarTse00}, and Durisi et al. \cite{DurisiBoelcskeiShamai06}  demonstrate that the capacity of such channels, in the wideband limit, is the same as for a wideband additive Gaussian noise channel with no fading, but the input signals, such as $M$-ary FSK, are highly bursty in the frequency domain or time domain.  The work of
Medard and Gallager \cite{MedardGallager02} (also see \cite{SubramanianHajek02}) shows that if the burstiness of the input signals is limited in both time and frequency, then the capacity of such wideband channels becomes severely limited.  In particular, the required energy per bit converges to infinity. 

Wireless wideband channels typically include both time and frequency selective fading. One approach to modeling such channels is to partition the frequency band into narrow subbands, so that the fading is flat, but time-varying, within each subband. If the width of the subbands is approximately the coherent bandwidth of the channel, then they will experience approximately independent fading. The flat fading models used in this paper can be considered to be models for communication over a subband of a wideband wireless fading channel.  The peak-power constraints that we impose on the signals can then be viewed as burstiness constraints in both the time and frequency domain for wideband communication, similar to those of \cite{MedardGallager02,SubramanianHajek02}. However, in this paper, we consider hard peak constraints, rather than fourth moment constraints as in \cite{MedardGallager02,SubramanianHajek02}, and we consider the use of multiple antennas.

The recent work of Srinivasan and Varanasi \cite{SrinivasanVaranasi06} is closely related
to this paper.   It gives low SNR asymptotics of the capacity of MIMO channels with no side information for block fading channels, with peak and average-power constraints,
with the peak constraints being imposed on individual antennas.
One difference between \cite{SrinivasanVaranasi06} and this paper is that
we assume continuous fading rather than block fading. 
In addition, we provide upper bounds on capacity rather than only asymptotic bounds
as in \cite{RaoHassibi04,SrinivasanVaranasi06}.   We assume, however, that the
fading processes are Rayleigh distributed, whereas the asymptotic bounds do not
require such distributional assumption.
The work of Rao and Hassibi \cite{RaoHassibi04} is also related to this paper.
 It gives low SNR asymptotics of the capacity of MIMO channels with no side
information for block fading channels, but the peak constraints are imposed on
coefficients in a particular signal representation, rather than as hard constraints
on the transmitted signals.

The model in this paper considers both a peak constraint and an average power constraint.  Upper bounds are given on the capacity which are valid for any ratio of these constraints, but the low SNR asymptotics focuses only on the case where the ratio is constant.  The ratio is also held constant in the asymptotic analysis of Srinivasan and Varinasi \cite{SrinivasanVaranasi06}.  The paper of Wu and Srikant \cite{WuSrikant07} focuses on the asymptotic capacity and error exponent for a fixed peak constraint, as the average power goes to zero. The paper of Zheng et. al  \cite{ZhengMedardTse} considers a general scaling of the peak constraint to average power constraint, with the scaling depending also on the coherence time.   For a fixed ratio of peak constraint to average power constraint, the capacity scales quadratically as SNR converges to zero, whereas for a fixed peak constraint, the capacity scales linearly with capacity as SNR converges to zero.   Cases between these two extremes are investigated in  \cite{ZhengMedardTse}.  For wideband cellular systems using OFDM modulation, the peak constraint is usually expressed in the time domain, because of the limitations on the linear range of transmit power amplifiers.   In such case, the peak power constraint in a particular frequency is not severe, so letting the peak constraint be constant or letting it converge to zero more slowly than the average power may be most appropriate.   In cases in which interference with other users within the same band is especially important, for example for use of unlicensed or secondary spectrum, a peak constraint of the same order of magnitude as the average power constraint, as considered in this paper, may be the most relevant.  The papers \cite{SrinivasanVaranasi06,WuSrikant07,ZhengMedardTse} consider block fading channels, whereas a stationary, correlated fading channel model is adopted here.

The capacity of noncoherent stationary flat fading channels at high SNR was studied in \cite{LapidothMoser,Lapidoth03,Lapidoth05,KochLapidoth}, and the capacity of delay spread channels at high SNR was recently studied in \cite{KochLapidoth08}. For regular fading processes \cite{LapidothMoser} demonstrated a connection between the high-SNR capacity growth and the error in predicting the fading process from \emph{noiseless} observations of its past, whereas for
nonregular fading \cite{Lapidoth05} demonstrated such a connection to the error in predicting the fading process from \emph{noisy} observations of its past in the low observation noise regime. In this paper we point to an analogous connection between the low SNR asymptotic capacity and the error in predicting the fading process from \emph{very noisy} observations of its past. We show that these prediction errors in the high observation noise regime determine the asymptotic low SNR capacity of SISO channels and MIMO channels with sum power constraints. They also determine the capacity of a class of MIMO channels satisfying a certain separability condition.
Our results on delay spread channels follow from those on MIMO channels with individual power constraints. 

The rest of this paper is arranged as follows. In Section \ref{sec:mainresults} we describe the channel models that are considered in this paper and present the main capacity results obtained with these channel models. In Sections \ref{sec:SISO} to \ref{sec:delay} the capacity results are proved--in some
cases by exhibiting additional capacity bounds. 


\section{Channel Models and Main Results}\label{sec:mainresults}

We study four types of channels: SISO channels, MIMO channels with sum (across transmit antennas) power constraints, MIMO channels with individual (per transmit antenna) power constraints, and
SISO channels with delay spread. In this section we shall describe
these models and present our results on their capacities.


\subsection{SISO channels}\label{sub:SISO}

We begin with the SISO channel, which models noncoherent
discrete-time single-antenna communication over time-selective flat
fading channels.

\subsubsection{The Model}

The time-$k$ complex-valued output $Y_k \in \Complex$ of the SISO
channel is given by
\begin{equation}\label{eq:SISOmodel}
        Y_k = \sqrt{\rho} H_k z_k + W_k,
\end{equation}
where $z_{k} \in \Complex$ is the time-$k$ channel input; the SNR $\rho$
is a positive scaling constant;
the complex stochastic process $\{ H_k\}$ is the multiplicative fading process
and the complex stochastic process $\{ W_k \}$ models additive noise.

We assume that the processes $\{H_k\}$ and $\{ W_k \}$ are independent
and that their joint law does not depend on the input sequence
$\{z_{k}\}$.  The additive noise sequence $\{W_k\}$ is a sequence of
IID proper complex normal (PCN) random variables of mean zero and
variance one.  Such a distribution is denoted by $\NormalC{0}{1}$. The
fading process $\{H_k\}$ is assumed to be a zero-mean, unit-variance,
stationary, PCN process. We denote its autocorrelation function by
$R(\cdot)$ and assume that it has a spectral density function
$S(\cdot)$. Thus
\begin{align*}
        R(m) & \triangleq \E {H_{k+m} H_k^*} \\
        & =  \int _{-\pi}^\pi S(\omega) e^{i m \omega}\frac{\d \omega}{2\pi}, 
        \qquad k,m\in\Integers
\end{align*}
and, in particular,
\begin{equation*}
        R(0) = \E { |H_k|^2 } = 
        \int_{-\pi}^{\pi} S(\omega)\frac{\d \omega}{2\pi} = 1, \qquad k \in \Integers.
\end{equation*}
Note that the existence of its spectral density function implies that $\{H_k\}$ is ergodic. We shall assume throughout that the autocorrelation is
square-summable, i.e., that
\begin{equation}
  \label{eq:square-summable}
  \sum_{\nu=-\infty}^{\infty} |R(\nu)|^{2} < \infty
\end{equation}
and define
\begin{align}
  \label{eq:def_lambda}
  \lambda & \triangleq \sum_{\nu=-\infty}^\infty |R(\nu)|^2 \\
  & = \int_{-\pi}^\pi |S(\omega)|^2 \frac{\d \omega}{2\pi} \nonumber.
\end{align}

The input is simultaneously subjected to two power constraints: a peak power constraint and an
average power constraint. The peak power constraint is that
the time-$k$ channel input $Z_{k}$ must satisfy, with probability one, 
\begin{equation}\label{eq:SISOpeak}
        |Z_k| \le 1, \qquad k \in \Integers.
\end{equation}
The average-power constraint is that
\begin{equation}
  \label{eq:SISOaverage}
        \E {|Z_k|^2} \le \frac{1}{\beta}, \qquad k \in \Integers,
\end{equation}
where the peak-to-average ratio $\beta$ is some constant
satisfying $\beta \geq 1$ and is the ratio of the maximum allowed peak
power to the maximum allowed average power. Since the average of a
random variable cannot exceed its maximal value, it follows that \eqref{eq:SISOpeak} implies $\E{|Z_k|^2} \le 1$, so that
setting $\beta=1$ renders the average power constraint inactive and
thus reduces the problem to one of communication subject to a
peak-power constraint only.

The capacity of this channel is given by
\begin{equation*}
        C(\rho, \beta) = \lim_{n\to\infty} \frac{1}{n} \sup I(Z_1^n; Y_1^n),
\end{equation*}
where the supremum is taken over all joint distributions on $Z_1^n$
satisfying the peak power constraint \eqref{eq:SISOpeak} and the average
power constraint \eqref{eq:SISOaverage}. 
The square-summability assumption \eqref{eq:square-summable}, together with the assumption
that  $\{H_k\}$ is PCN, implies that the random process
$\{H_k\}$ is weakly mixing (in fact, mixing).  Therefore, a coding theorem exists
for  $C(\rho,\beta)$, based on  notions surrounding information stability (see \cite{Pinsker,Gray})
and the  Shannon--McMillan--Breiman theorem for finite-alphabet 
ergodic sources.   See \cite{VigneshHajek05} for details.
Roughly speaking,  the operational meaning of $C(\rho,\beta)$ is that for any rate $R$
(expressed in bits per channel use)  less than the capacity,  there exists
a sequence of codes with blocklength converging to infinity,
such that each code meets peak and average power constraints, each code has
 $2^{nR}$ codewords, where $n$ is the length of the code, and the probability of
decoding any codeword incorrectly converges to zero.   The average power
constraint can be imposed on the expectation (over a uniformly chosen codeword) or on
the maximum (over all the codewords) of the normalized (by the blocklength)
energy of the codeword.

We define $c(\beta)$ as the limiting ratio
\begin{equation}\label{eq:c_beta}
        c(\beta) \triangleq \lim_{\rho \downarrow 0} \frac{C(\rho, \beta)}{\rho^2}
\end{equation}
when the limit exists. We next present our results
on $C(\rho,\beta)$.

\subsubsection{Results on SISO Fading}
Our first result gives the asymptotic capacity of the SISO channel.

\begin{proposition}[Asymptotic Capacity]
  \label{prop:SISOcapacity}
For any $\beta \geq 1$, the limit in \eqref{eq:c_beta} exists and is given by  
\begin{eqnarray}
  \label{eq:SISOcapacity}
  c(\beta) & = & \frac{1}{2} \cdot \max_{0\le a \le\frac{1}{\beta}} \left\{ a\lambda - a^2 \right\} \\
  \label{eq:SISOcapacity2}
  & = &
  \begin{cases}  
    \frac{\lambda^2}{8} &
    \text{if $\displaystyle{\lambda < \frac{2}{\beta}}$} \\
    \frac{\lambda}{2\beta} - \frac{1}{2\beta^2} & \text{if
      $\displaystyle{\lambda \ge \frac{2}{\beta}}$}
\end{cases}.
\end{eqnarray}
\end{proposition}

By evaluating the RHS of \eqref{eq:SISOcapacity} for the special case
where $\beta=1$, i.e., when the average power constraint is inactive,
we obtain
\begin{corollary}[Asymptotic Capacity---No Average Power Constraint]
  \label{coro:SISOnoaverage}
  Under   the    peak-power   constraint   \eqref{eq:SISOpeak}   only,
  \begin{equation}
    \label{eq:asy_peak_only}
    c(1) = 
    \begin{cases} 
      \frac{\lambda^2}{8} & \text{if $\lambda < 2$} \\ 
      \frac{\lambda-1}{2} & \text{if $\lambda \ge 2$}
    \end{cases}.
  \end{equation}
\end{corollary}

Motivated by the different asymptotic behaviors of channel capacity
that occur depending on whether $\lambda < 2$ or $\lambda \geq 2$, we
introduce the following definition.
\begin{definition}
A zero-mean discrete-time PCN stationary process $\{H_{k}\}$ (not necessarily of unit
  variance) is {\em ephemeral} if its autocorrelation function
  $R(\cdot)$ satisfies
  \begin{equation}
    \sum_{\nu=-\infty}^{\infty} |R(\nu)|^{2} < 2 R^2 (0).
  \end{equation}
  Otherwise, $\{H_{k}\}$ is {\em nonephemeral}.
\end{definition}
Note that if the fading process $\{H_{k}\}$ is of unit-variance, then
$R(0)=1$ and $\{H_{k}\}$ is ephemeral if $\lambda < 2$, where
$\lambda$ is defined in \eqref{eq:def_lambda}.
When the fading process in the SISO fading channel \eqref{eq:SISOmodel}
is ephemeral we consider the channel itself to also be ephemeral.
Otherwise, we consider the channel to be nonephemeral.

In addition to asymptotic expansions, we provide a firm upper
bound on $C(\rho, \beta)$:
\begin{proposition}[A Firm Upper Bound on Capacity]
  \label{prop:SISOupper}
  For any $\rho>0$ and $\beta\ge 1$,
  \begin{equation}
    \label{eq:SISOupper}
    C(\rho, \beta) \le U(\rho, \beta),
  \end{equation}
where
\begin{equation}
  \label{eq:def_U}
U(\rho, \beta) \triangleq \log \bigl( 1+\rho \zeta(\rho, \beta) \bigr) - 
\zeta(\rho, \beta) I(\rho),
\end{equation}
\begin{equation}\label{eq:def_zeta}
  \zeta(\rho,\beta) \triangleq \min \left\{ 
    \frac{1}{\beta}, \frac{1}{I(\rho)}-\frac{1}{\rho} 
  \right\},
\end{equation}
and
\begin{equation}
  \label{eq:def_I}
  I(\rho)  \triangleq \int_{-\pi}^{\pi} 
  \log \bigl( 1+\rho S(\omega) \bigr) \frac{\d \omega}{2 \pi}.
\end{equation}
\end{proposition}

It is interesting to note that, in general, IID input distributions do
not achieve the same asymptotic behavior as channel capacity. This is
best seen in the next proposition on the asymptotic behavior of the
mutual information corresponding to IID inputs. We first define
\begin{equation}\label{eq:c_iid_beta}
        c_{\text{IID}}(\beta) \triangleq \lim_{\rho \downarrow 0} \frac{1}{\rho^{2}} 
    \left(  \lim_{n \rightarrow \infty} \sup \frac{1}{n} I(Z_1^n; Y_1^n) \right)
\end{equation}
if the limit exists, where the supremum is over all IID distributions on $Z_{1}^{n}$ satisfying \eqref{eq:SISOpeak} and \eqref{eq:SISOaverage}.

\begin{figure}[h]
        \centering
        \psfrag{c}{\small$c(\beta)$}
        \psfrag{ciidciid}{\small$c_{\textnormal{IID}}(\beta)$}
        \psfrag{lambda}[cc][cc]{$\lambda$}
        \psfrag{Asymptotic Capacity}{\small Asymptotic Capacity}
        \psfrag{0}[rr][rr]{\small$0$}
        \psfrag{1}[rr][rr]{\small$1$}
        \psfrag{2}[rr][rr]{\small$2$}
        \psfrag{3}[rr][rr]{\small$3$}
        \psfrag{4}[rr][rr]{\small$4$}
        \psfrag{5}[rr][rr]{\small$5$}
        \psfrag{0.5}[rr][rr]{\small$0.5$}
        \psfrag{1.5}[rr][rr]{\small$1.5$}
        \includegraphics[width=0.6\textwidth]{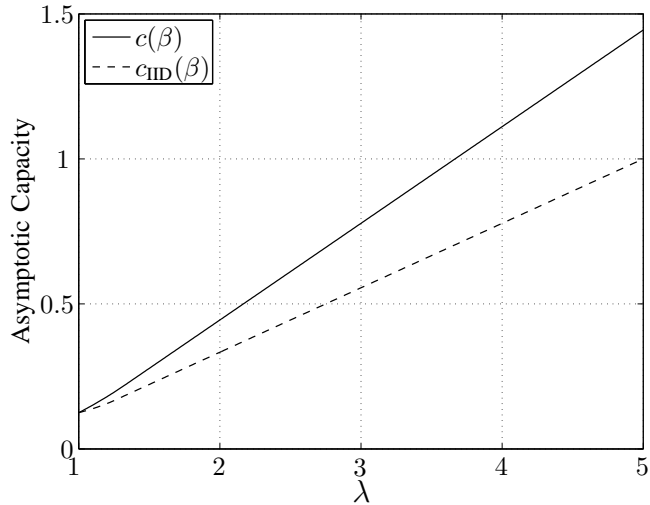}
        \caption{Comparison of $c(1.5)$ (asymptotic capacity) and $c_{\text{IID}}(1.5)$ (asymptotic information rates achievable with IID input symbols). Peak-to-average ratio $\beta=1.5$.} \label{fig:iid}
\end{figure}

\begin{proposition}[Asymptotic Rates for IID Inputs]
  \label{prop:SISOiid}
  If the autocorrelation function $R(\cdot)$ is absolutely summable,
  i.e.,
  \begin{equation}
    \label{eq:summable}
    \sum_{\nu=-\infty}^{\infty} |R(\nu)| < \infty,
  \end{equation}
  then the limit in \eqref{eq:c_iid_beta} exists and is given by
  \begin{equation}
    c_{\textnormal{IID}}(\beta) = 
    \begin{cases} 
      \frac{1}{8(2-\lambda)} & \textnormal{if 
        $\displaystyle{\lambda < 2 - \frac{\beta}{2}}$} \\ 
      \frac{1}{2\beta} + \frac{\lambda-2}{2\beta^2} & \text{if 
        $\displaystyle{\lambda \ge 2 - \frac{\beta}{2}}$}
    \end{cases}.
  \end{equation}
\end{proposition}
Using this proposition we see that, subject to \eqref{eq:summable}
(which is more stringent than \eqref{eq:square-summable}), IID inputs
achieve the asymptotic behavior of channel capacity only if $\lambda =
1$ (in which case the channel is memoryless) or when the two
conditions $\beta=1$ and $\lambda \geq 2$ are both met.
Figure~\ref{fig:iid} depicts $c(1.5)$ and $c_{\textnormal{IID}}(1.5)$ as functions of
$\lambda$.

\subsubsection{Discussion}

A few remarks about the results are called for. 
\begin{itemize}
\item \emph{Capacity and Prediction}: The error in predicting the time-zero fading $H_{0}$ based on the previous values
  of the fading $\{H_{\nu}\}_{-\infty}^{-1}$ was shown in
  \cite{LapidothMoser} to be related to the high SNR asymptotic
  behavior of channel capacity. If this prediction error is strictly
  positive, then capacity at high SNR grows double logarithmically in
  the SNR. In cases where this prediction error is zero, a finer
  analysis of the prediction problem is needed to establish the high
  SNR capacity asymptotics \cite{Lapidoth05}. Indeed, when the
  prediction error is zero, the capacity asymptotics are determined
  by the behavior of the \emph{noisy} prediction error. This noisy
  prediction error $\sigma^2(\rho)$ is defined as the mean squared-error 
  in predicting $H_0$ based on $\left(\ldots, H_{-2}+N_{-2},
    H_{-1}+N_{-1} \right)$ where $\{N_k\}$ are IID
  $\NormalC{0}{1/\rho}$ and independent of $\{H_k\}$. 
  Furthermore, $\sigma^2(0)$ is assigned the value $R(0)$ corresponding to the limiting case of estimating $H_0$ in the absence of past information. 
  It is given by classical
  formulas for optimal prediction of stationary random processes by (see  \cite{Lapidoth05} for
  details):
  \begin{equation}\label{eq:sigma}
    \sigma^2(\rho) = \exp \left\{ \int_{-\pi}^{\pi} \log 
      \left( \frac{1}{\rho} + S(\omega) \right) \frac{\d \omega}{2\pi}
    \right\} - 
    \frac{1}{\rho}, \quad \rho > 0 .
  \end{equation}
  Here we note that the noisy prediction error also determines the asymptotic behavior of channel
  capacity at low SNR. Indeed, by Proposition~\ref{prop:SISOcapacity}, the low SNR
  asymptotics are determined by $\lambda$, which is defined in
  \eqref{eq:def_lambda}, and is related to the
  behavior of the noisy prediction error in the following way: 
  the Taylors series expansion of $\sigma^2(\rho)$ is
  \begin{equation} \label{eq:sigma_lambda}
  \sigma^2(\rho) = R(0) - \frac{\lambda-R^2(0)}{2}\rho + o(\rho),
  \end{equation}
where the notation $o(\cdot)$ is used in the sense that $\lim_{x\to 0}
o(x)/x = 0$. We further note that $I(\rho)$, defined in \eqref{eq:def_I}, can be expressed as
\begin{equation}
  \label{eq:I_sigma}
  I(\rho) = \log \left(1+\rho\sigma^2(\rho)\right),
\end{equation}
and \eqref{eq:sigma_lambda} is equivalent to
\begin{equation}
  \label{eq:I_rt_rho}
  I(\rho) = R(0)\cdot \rho - \frac{\lambda \rho^2}{2} + o(\rho^2).
\end{equation}
For fading processes that have unit variance, \eqref{eq:sigma_lambda} and \eqref{eq:I_sigma} still hold, while \eqref{eq:I_rt_rho} becomes
\begin{equation}
  \label{eq:IrhoSISO}
  I(\rho) = \rho - \frac{\lambda \rho^2}{2} + o(\rho^2).
\end{equation}

The proof of \eqref{eq:sigma_lambda}, \eqref{eq:I_rt_rho} and \eqref{eq:IrhoSISO} is a straightforward
application of the second order Taylor expansion of the function $x \mapsto \log(1+x),$
\begin{equation*}
        \log(1+x) = x - \frac{x^2}{2} + o(x^2),
\end{equation*}
and the monotone convergence theorem.
\item \emph{Input Distributions}: The proof of the achievability part of
  Proposition~\ref{prop:SISOcapacity} demonstrates that the low SNR
  asymptotics of channel capacity can be achieved by considering joint
  distributions on $Z_{1}, \ldots, Z_{n}$ of the form
  \begin{equation}
    \label{eq:tram10}
    Z_k = U \cdot \Phi_k, \quad 1 \leq k \leq n,
  \end{equation}
  where $U$ is a random variable taking value in $\{0,1\}$ and where
  the sequence $\{\Phi_k\}$ is independent of $U$ and consists of
  zero-mean modulus-1 random variables that are uncorrelated. 
  The amplitude modulation component of the optimal signaling strategy is captured in the law of $U$, and the phase  modulation component is captured in the law of $\Phi$. 
  Some examples of distributions on $\{ \Phi_k\}$ are the following:
\begin{itemize}
        \item[i)] $\Phi_k = \exp( i \cdot k \Theta)$, where $i=\sqrt{-1}$, and where $\Theta$ is a discrete random variable uniformly distributed over the set $\left\{ \frac{j \cdot 2\pi}{m} : j \in \{0,\ldots,m-1\}\right\}$ for some integer $m>1$. This is $m$-ary frequency shift keying (FSK);
        \item[ii)] $\{ \Phi_k \}$ are IID random variables uniform over the set $\left\{ \exp(i \cdot \frac{j\cdot 2\pi}{m}) : j \in\{0,\ldots, m-1\} \right\}$ for some integer $m>1$. This is $m$-ary phase shift keying (PSK);
        \item[iii)] $\{ \Phi_k \}$ are IID random variables uniformly distributed over the set $\{ \exp (i \theta): \theta \in [0,2\pi) \}$. This is also a form of PSK.
\end{itemize}
  In practice,  the signal of duration $n$ described in \eqref{eq:tram10} would be considered
as a single symbol, and, as is usual in the theory of channel coding, longer random
codewords would be comprised of many independent length $n$ symbols.
 Note that even when $\{ \Phi_k \}$ are
  IID (as when PSK is used), the random variables $\{ Z_k \}$ need not
  be IID because $\{ Z_k \}$ all have the same magnitude (namely,
  $U$). Thus, whenever $U$ is not deterministic, the sequence $\{ Z_k
  \}$ is not IID.  The fact that our proposed input distribution \eqref{eq:tram10} does
  not render $\{Z_{k}\}$ IID should not be surprising because IID
  inputs do not typically achieve the low SNR channel capacity
  asymptotics; see Proposition~\ref{prop:SISOiid}. 
  When the fading channel is not memoryless, then IID inputs achieve the asymptotic
  capacity only if there is no average power constraint and if there is sufficient memory in the channel ($\lambda \geq 2$.) Also, when there is sufficient memory in the channel, amplitude modulation (nondeterministic $U$) is needed whenever the average and peak power constraints differ. The ON-OFF ratio of $U$ is then determined by the ratio of the average to peak power constraints.  
  \item \emph{Relation to STORM}: 
  The FSK version (i.e. case (i) above) of the
  input distribution  \eqref{eq:tram10}  is the single antenna special case of
  the space time orthogonal rank one modulation (STORM)
  input distribution, proposed for  MIMO block fading channels by 
  Srinivasan and Varanasi \cite{SrinivasanVaranasi06}.   The distribution is
  used differently here. 
  In  \cite{SrinivasanVaranasi06} the parameter
  $n$ is taken to equal the block length of the channel.   For the stationary fading
  considered here, the capacity asymptote is approached by letting $n\rightarrow\infty.$

\item \emph{PSK Inputs}: Zhang \& Laneman \cite{Laneman05} studied the
  low-SNR asymptotic behavior of the information rates that can be
  achieved on our channel when PSK inputs are used. In the language of
  \eqref{eq:tram10}, PSK inputs correspond to choosing $U$ in
  \eqref{eq:tram10} to be deterministically equal to one and
  $\{\Phi_{k}\}$ to be a sequence of PSK input symbols (as described in ii) or iii) above). (The constellation of the PSK does not affect the asymptotic information rate.) The asymptotic
  rates achieved by PSK (with $\beta = 1$) were derived in
  \cite{Laneman05} and are given by
  \begin{equation}
    \label{eq:Laneman}
    c_{\textnormal{PSK}} \triangleq \lim_{\rho \downarrow 0} \frac{C_{\textnormal{PSK}}(\rho)}{\rho^{2}}
    = \frac{\lambda - 1}{2}, \quad (\beta
    = 1),
  \end{equation}
  where $C_{\text{PSK}}(\rho)$ denotes the information rate achieved using PSK inputs.
  PSK is, in general, suboptimal when $\beta > 1$ because in PSK the
  peak power and the average power are the same. Even when $\beta=1$,
  PSK is not always optimal. It is optimal for nonephemeral channels
  because for $\beta = 1$ and $\lambda \geq 2$ the RHS of
  \eqref{eq:Laneman} agrees with the RHS of \eqref{eq:asy_peak_only}:
\begin{equation}
 c_{\textnormal{PSK}} = c(1), \quad (\lambda \geq 2, \ \beta=1).
\end{equation}
For $1.5 \leq \lambda < 2$,  PSK is only optimal among IID input
distributions
\begin{equation}
        c_{\textnormal{PSK}} = c_{\textnormal{IID}}(1) < c(1), \quad (1.5 \leq \lambda < 2, \ \beta=1).
\end{equation}
And for $1 \leq \lambda < 1.5$, PSK is not optimal even among IID input
distributions
\begin{equation}
  c_{\textnormal{PSK}} < c_{\textnormal{IID}}(1) < c(1), \quad (1 \leq \lambda < 1.5, \ \beta=1).
\end{equation}
Consider the special case when the channel is memoryless ($\lambda=1$): here, PSK does not
achieve any positive rates because it encodes all information in the phase of the transmit signal; the memoryless channel completely wipes this information out by adding a phase term that takes new independent realizations with time.  Unlike PSK, IID inputs that use amplitude modulation can achieve positive rates on the memoryless fading channel. Figure \ref{fig:psk} compares $c(1)$, $c_{\textnormal{IID}}(1)$, and $c_{\text{PSK}}$ as functions of $\lambda$.
\begin{figure}[h]
    \centering
    \psfrag{c}{\small$c(1)$}
    \psfrag{ciidciid}{\small$c_{\textnormal{IID}}(1)$}
    \psfrag{cpsk}{\small$c_{\textnormal{PSK}}$}
    \psfrag{lambda}[cc][cc]{$\lambda$}
    \psfrag{Asymptotic Capacity}{\small Asymptotic Capacity}
        \psfrag{0}[rr][rr]{\small$0$}
        \psfrag{0.2}[rr][rr]{\small$0.2$}
        \psfrag{0.4}[rr][rr]{\small$0.4$}
        \psfrag{0.6}[rr][rr]{\small$0.6$}
        \psfrag{0.8}[rr][rr]{\small$0.8$}
        \psfrag{1}[rr][rr]{\small$1$}
        \psfrag{1.5}[rr][rr]{\small$1.5$}
        \psfrag{2}[rr][rr]{\small$2$}
        \psfrag{2.5}[rr][rr]{\small$2.5$}
    \includegraphics[width=0.6\textwidth]{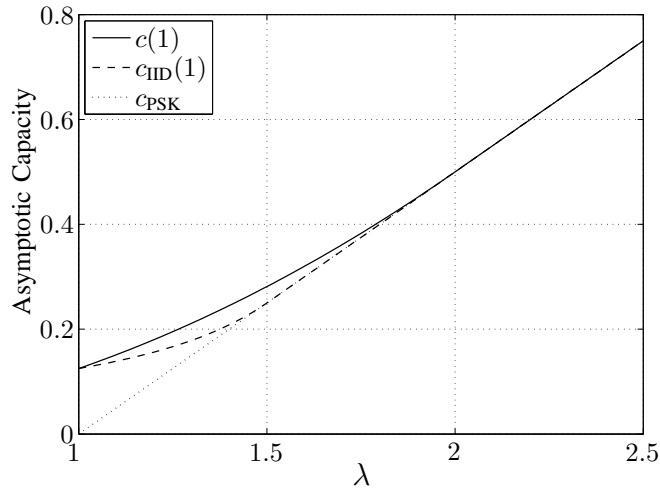}
    \caption{Comparison of $c(1)$, $c_{\text{IID}}(1)$ and $c_{\text{PSK}}$ (no average power constraint).}
  \label{fig:psk}
\end{figure}
\item \emph{On the Nonasymptotic Bound}: The nonasymptotic bound
  presented in Proposition~\ref{prop:SISOupper} is tight at low SNR in
  the sense that for any fixed $\beta \geq 1$ we have
  \begin{equation}
    \lim_{\rho \downarrow 0} \frac{U(\rho,\beta) }{\rho^{2}} =
    c(\beta).
  \end{equation}
It is also tight when $\rho > 0$ is held fixed and $\beta$ goes to
infinity in the sense that
\begin{equation}
  \label{eq:tram20}
\lim_{\beta\to\infty} \frac{\beta}{\rho}U(\rho, \beta) = \lim_{\beta
  \to \infty} \frac{\beta}{\rho} C(\rho, \beta), \quad \rho > 0,
\end{equation}
where the RHS of the above is given by \cite{VigneshHajek05}
\begin{equation}
  \label{eq:tram30}
\lim_{\beta
  \to \infty} \frac{\beta}{\rho} C(\rho, \beta) = 1 -
  \frac{I(\rho)}{\rho}, \quad \rho > 0.
\end{equation}
Thus, our upper bound could be used as an alternate to the upper bound
used in \cite{VigneshHajek05}.  Note that in fixing $\rho$ and letting
$\beta$ go to infinity we are holding the peak power fixed and
letting the allowed average power go to zero.

To verify \eqref{eq:tram20} one can compute the LHS of
\eqref{eq:tram20} and then show that it equals the RHS of
\eqref{eq:tram30}. This can be done by noting that for $\beta$
sufficiently large we have $\zeta(\rho, \beta) = 1/\beta$, and thus
\begin{equation*}
        U(\rho,\beta) = \log\left(1+\frac{\rho}{\beta} \right) - \frac{I(\rho)}{\beta} = \frac{\rho - I(\rho)}{\beta} + o(\beta^{-1}), \quad \beta \gg 1.
\end{equation*}

The upper bound $U(\rho,\beta)$ is found to be close to the channel capacity at nontrivial values of SNR. As a demonstration, we numerically compare the upper bound to the following lower bound derived in \cite{VigneshBH05}. 
Let \begin{equation}
L(\rho) = I(Z_0; Y_0 | Z_{-\infty}^{-1}, Y_{-\infty}^{-1})
	\label{eq:LfromVigneshBH06}
\end{equation}
where the input $\{Z_k\}$ is an IID Quadrature PSK process. The channel capacity satisfies
\begin{equation}
  C(\rho,\beta) \geq \max_{\gamma \in [1,\beta]} \frac 1 {\gamma} L(\frac{ \gamma \rho}{ \beta} ) .
  \label{eq:LBfromVigneshBH06} 
\end{equation} 
In Figure \ref{fig:Beta10}
, we graph the capacity bounds $U$ and $L$ for a Gauss-Markov channel with correlation coefficient $0.99$. Numerical integration is used to compute the lower bound. The peak-to-average ratio is set as $10$. The bounds are found to be fairly tight.
\begin{figure}[h]
    \centering
    	\includegraphics[width=0.6\textwidth]{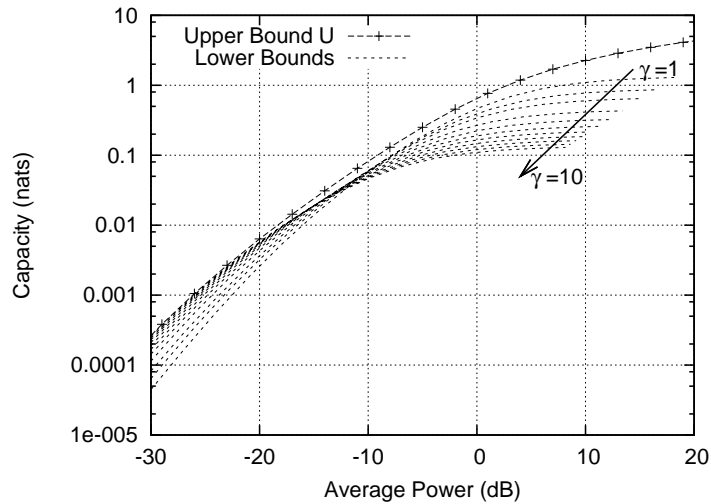}
        \caption{Comparison of upper and lower bounds on capacity. Gauss-Markov channel with correlation coefficient $0.99$. Peak-to-average ratio $\beta=10$.} \label{fig:Beta10}
\end{figure}
\end{itemize}


\subsection{MIMO channels with sum power constraints}\label{sub:MIMOsum}

For MIMO channels, we separately consider two different types of
constraints on the input signal. The constraints are imposed either on
sums across transmit antennas, or on individual transmit antennas.
This subsection is devoted to MIMO channels with sum power
constraints.

\subsubsection{The Model\label{subsub:MIMOmodel}} 
We consider a single user discrete-time MIMO channel with $\nt$
transmit antennas and $\nr$ receive antennas. The time-$k$ channel
output $\vect{Y}_k \in \Complex^{\nr}$ is given by
\begin{equation}\label{eq:MIMOmodel}
        \vect{Y}_k = \sqrt{\rho} \vmat{H}_k \vect{z}_k + \vect{W}_k,
\end{equation}
where $\sqrt{\rho} \vect{z}_k$ is the time-$k$ channel input, with
$\rho>0$ representing the peak SNR and $\vect{z}_k \in
\Complex^{\nt}$. In the above, the multiplicative noise $\{\vmat{H}_k
\}_{k=-\infty}^{\infty}$ is a matrix-valued stochastic process such
that at every time instant $k \in \Integers$, the random matrix
$\vmat{H}_k$ is an $\nr \times \nt$ complex random matrix. The random
vectors $\{\vect{W}_k\}_{k=-\infty}^{\infty}$ are IID random vectors,
each consisting of $\nr$ independent $\NormalC{0}{1}$ components.
Thus, $\vect{W} \sim \NormalC{0}{\mat{I}_{\nr}}$, where
$\mat{I}_{\nr}$ is the $\nr\times\nr$ identity matrix.  Denoting by
$H_k^{(r,t)}$ the row-$r$ column-$t$ entry in $\vmat{H}_k$, we can
write the $r$-th element in $\vect{Y}_k$ as
\begin{equation}\label{eq:MIMOmodel1}
        Y_k^{(r)} = 
        \sqrt{\rho} \sum_{t=1}^{\nt} H_k^{(r,t)} z_k^{(t)} + W_k^{(r)}.
\end{equation}

As for SISO channels, we assume that $\{\vmat{H}_k\}$ and
$\{\vect{W}_k\}$ are independent, and that their joint law does not
depend on $\{\vect{Z}_k\}$.  We further assume that for each pair
$(r,t)$ satisfying
\begin{equation}
  \label{eq:range}
  (r,t) \in \{1, \ldots, \nr\} \times \{1, \ldots, \nt\}
\end{equation}
the scalar process $\{ H_k^{(r,t)}\}_{k=-\infty}^{\infty}$ is a
zero-mean, stationary, PCN process with autocorrelation function
$R_{r,t}(\cdot)$ and spectral density function $S_{r,t}(\cdot)$. We
also assume that the $\nr \times \nt$ processes corresponding to the
different pairs $(r,t)$ satisfying \eqref{eq:range} are independent.
We finally assume throughout this paper that autocorrelation
$R_{r,t}(\cdot)$ is square-summable for every antenna pair $(r,t)$
satisfying \eqref{eq:range} and define
\begin{align}
  \label{eq:def_lambda_rt}
  \lambda_{r,t} & \triangleq \sum_{\nu=-\infty}^\infty |R_{r,t}(\nu)|^2 \\
  & = \int_{-\pi}^\pi |S_{r,t}(\omega)|^2 \frac{\d \omega}{2\pi} \nonumber.
\end{align}

\begin{definition}
  A MIMO channel is said to be {\em nonephemeral} if for every
  pair $(r,t)$ satisfying \eqref{eq:range}, the fading process
  $\{H_k^{(r,t)}\}$ is nonephemeral, i.e., if 
  \begin{equation}
   \lambda_{r,t} \ge 2 R_{r,t}^2(0), \quad \text{for all} \quad  
   (r,t) \in \{1, \ldots, \nr\} \times \{1, \ldots, \nt\}.
  \end{equation}
\end{definition}

\begin{definition}\label{def:transmit_separable}
  A MIMO channel is said to be {\em transmit separable} if there
  are $\nt$ nonnegative constants $\left\{ \alpha_t :
    t\in\{1,\ldots,\nt\} \right\}$ and $\nr$ autocorrelation functions
  $\left\{R_r(\cdot) : r \in \{1,\ldots,\nr\} \right\}$ with
  corresponding spectral density functions $S_r(\cdot)$ such that
  \begin{equation*}
  	R_{r,t}(k) = \alpha_t R_r (k)
  \end{equation*}
  for all $ ( r, t ) \in \{1,\ldots,\nr\}\times\{1,\ldots,\nt\}$ and $k\in \Integers$.
\end{definition}

Definition \ref{def:transmit_separable} says that a MIMO channel is
transmit separable if fixing any one receive antenna, the channels
from all the transmit antennas have the same law up to some scaling
constants.

\subsubsection{The sum power constraints} 
The {\em sum peak-power constraint} on the channel inputs is that
the time-$k$ channel input $Z_k$ must satisfy, with probability one, 
\begin{equation}\label{eq:sumpeak}
        \norm{\vect{Z}_k}_2 \le 1 , \quad k \in \Integers,
\end{equation}
where $\norm{\vect{Z}_k}_2^2$ denotes the sum of the squares of the components of $\vect{Z}_k$.
The {\em sum average-power constraint} is that
\begin{equation}\label{eq:sumaverage}
        \E{\norm{\vect{Z}_k}_2^2} \le \frac{1}{\beta}, \quad k \in \Integers.
\end{equation}
The capacity of the channel under the sum power constraints \eqref{eq:sumpeak} and \eqref{eq:sumaverage} is denoted by $C_{\text{S}}(\rho, \beta) $ and is given by
\begin{equation}
        C_{\text{S}}(\rho, \beta) = \lim_{n\to\infty} \frac{1}{n} \sup I(\vect{Z}_1^n; \vect{Y}_1^n)
\end{equation}
with the supremum taken over all distributions on $\vect{Z}_1^n$ satisfying \eqref{eq:sumpeak} and \eqref{eq:sumaverage}. We further define
\begin{equation}\label{eq:c_s_beta}
        c_{\text{S}}(\beta) \triangleq \lim_{\rho \downarrow 0} \frac{C_{\text{S}}(\rho, \beta)}{\rho^2},
\end{equation}
when the limit exists.

\subsubsection{Results on MIMO with sum power constraints}

The asymptotic low SNR capacity of the MIMO channel under sum power constraints is given in the following proposition.

\begin{proposition}[Asymptotic Capacity]\label{prop:MIMOScapacity}
        For any $\beta\ge 1$, the limit in \eqref{eq:c_s_beta} exists and is given by
        \begin{eqnarray}\label{eq:MIMOScapacity}
                c_{\text{S}}(\beta)  = \frac{1}{2} \sup_{\vect{a} \in \set{A}(\beta)} \sum_{r=1}^{\nr} \Bigg\{ \sum_{t=1}^{\nt} a_t \lambda_{r,t} ~~~~~~~\nonumber \\  - \left(\sum_{t=1}^{\nt} a_t R_{r,t}(0) \right)^2 \Bigg\}.
        \end{eqnarray}
        where
        \begin{equation}\label{eq:defAbeta}
                \set{A}(\beta) \triangleq \left\{ (a^{(1)},\ldots,a^{(\nt)}) \in \Reals^{\nt}: a_t\ge 0\ \forall\ t, \sum_{t=1}^{\nt} a_t\le \frac{1}{\beta} \right\}.
        \end{equation}
\end{proposition}

For transmit separable channels, the above proposition is simplified to the following corollary.

\begin{corollary}[Asymptotic Capacity---Transmit Separable]\label{coro:MIMOSseparable}
        If the MIMO channel is transmit separable, then
        \begin{equation}\label{eq:MIMOSseparable}
                c_{\text{S}}(\beta) = \frac{\alpha_{\max}^2}{2} \max_{0\le a \le \frac{1}{\beta}} \sum_{r=1}^{\nr} \left\{ a \lambda_r - a^2 R_r^2(0)\right\},
        \end{equation}
        where
        \begin{equation}\label{eq:def_alpha_max}
                \alpha_{\max} \triangleq \max \{ \alpha_1,\ldots,\alpha_{\nt} \},
        \end{equation}
        and for every $r$,
        \begin{align}\label{eq:def_lambda_r}
                \lambda_{r} & \triangleq \sum_{\nu=-\infty}^\infty |R_{r}(\nu)|^2 \\
  & = \int_{-\pi}^\pi |S_{r}(\omega)|^2 \frac{\d \omega}{2\pi} \nonumber.
        \end{align}
\end{corollary}

\begin{corollary}[Asymptotic Capacity---Transmit Separable, Nonephemeral, No Average Power Constraint]\label{coro:MIMOSnonephemeral}
        If the channel is transmit separable and nonephemeral, and if no average-power constraint is imposed ($\beta=1$), then
        \begin{equation}
          \label{eq:MIMOSnonephemeral}
     c_{\text{S}}(1) = 
     \alpha_{\max}^2 \sum_{r=1}^{\nr} \frac{\lambda_r - R_r^2(0)}{2},
        \end{equation}
        where $\lambda_r$ and $\alpha_{\max}$ are as defined in
        Corollary \ref{coro:MIMOSseparable}.
\end{corollary}

As for SISO channels, we also give a firm upper bound on $C_{\text{S}}(\rho,\beta)$.

\begin{proposition}[A Firm Upper Bound on Capacity of MIMO with Sum Constraints]\label{prop:MIMOSupper}
        For any $\rho>0$ and $\beta\ge 1$,
        \begin{equation}
                C_{\text{S}}(\rho, \beta) \le U_{\text{S}}(\rho, \beta),
        \end{equation}
        where
        \begin{eqnarray}
                U_{\text{S}}(\rho, \beta) \triangleq \max_{\vect{a}\in \set{A}(\beta)} \sum_{r=1}^{\nr} \left\{ \log\left(1+\rho\sum_{t=1}^{\nt} a_tR_{r,t}(0) \right) \right. \nonumber \\ \left. - \sum_{t=1}^{\nt} a_t I_{r,t}(\rho)\right\},~~~~~~~~\label{eq:MIMOSupper}
        \end{eqnarray}
        \begin{equation}
                I_{r,t}(\rho) \triangleq \int_{-\pi}^{\pi} \log\left(1+\rho S_{r,t}(\omega)\right)\frac{\d \omega}{2\pi},
        \end{equation}
        and $\set{A}(\beta)$ is defined as in \eqref{eq:defAbeta}.
\end{proposition}

\subsubsection{Discussion}
\begin{itemize}
\item \emph{Input Distributions:} As the proof of Proposition \ref{prop:MIMOScapacity} suggests, a distribution on $\vect{Z}_1^n$ that achieves the capacity asymptotically is the following. At most one of the $\nt$ transmit antennas is used during the whole transmission, with antenna $t$ being chosen with probability $a_t$. For the chosen antenna, all the input symbols $Z_1^{(t)}, \ldots, Z_n^{(t)}$ have magnitude one and their phases are chosen in such a way that each symbol is of mean zero and different symbols are uncorrelated. If no antenna is chosen, then all antennas keep silent during the whole transmission period.

        In the case when the MIMO channel is transmit separable (Corollary \ref{coro:MIMOSseparable}), the distribution on $\vect{Z}_1^n$ suggested in the proof is to only use the one strongest antenna (i.e., the $t$-th antenna with $\alpha_t = \alpha_{\max}$). The signals sent by this antenna have the same law as those used for SISO channels. 
        
        As for SISO channels, the suggested distributions for the above two cases (general and transmit separable) on $\vect{Z}_1^n$ are not IID. 

        Finally, for transmit separable, nonephemeral channels with no average-power constraint (Corollary \ref{coro:MIMOSnonephemeral}), the suggested input law is to use only the strongest antenna to send symbols that all have mean zero, magnitude one and that are uncorrelated in time.
        
        \item \emph{Comparison with SISO Channels:} We compare the asymptotic capacity of MIMO channels with sum power constraints to SISO channels. Consider the simple case when the MIMO channel satisfies
        \begin{equation} \label{eq:compareMIMOSISO}
                R_{r,t}(k) = R(k),\quad k \in \Integers,
        \end{equation}
        for every antenna pair $(r,t)$, where $R(\cdot)$ is the autocorrelation function of the SISO channel we compare the MIMO with. Note that such a MIMO channel is transmit separable. The asymptotic capacity of this MIMO channel follows from Corollary \ref{coro:MIMOSseparable}, and is given by
        \begin{equation*}
                c_{\text{S}}(\beta) = \frac{\nr}{2} \max_{0\le a \le \frac{1}{\beta}} \left\{ a \lambda - a^2\right\}  = \nr \cdot c(\beta).
        \end{equation*}
        Thus we see that the channel capacity at low SNR grows linearly with the number of receive antennas ($\nr$), but does not grow with the number of transmit antennas ($\nt$). The former observation is easy to understand, because the received signal energy is linear in $\nr$; the latter is not surprising when we recall that an optimal input distribution for MIMO channels with sum power constraints is to only use one transmit antenna at any time. Intuitively, having multiple transmit antennas is not helpful at low SNR because any benefit due to diversity brought by multiple transmit antennas is nulled by the cost of tracking the additional fading processes. 
        
\end{itemize}


\subsection{MIMO channels with individual power constraints}\label{sub:MIMOindiv}

The MIMO channel model we consider under individual power constraints is exactly the same as the model we consider under sum power constraints, as explained in \ref{subsub:MIMOmodel}. The difference is in the form of the power constraints.

\subsubsection{The individual power constraints}
The {\em individual peak-power constraint} on the MIMO channel is that
the time-$k$ channel input of the $t$-th antenna $Z_k^{(t)}$ must satisfy,  with probability one, 
\begin{equation}\label{eq:indivpeak}
        |Z_k^{(t)}| \le 1, \quad t \in \{ 1,\ldots, \nt \}, \ k\in\Integers.
\end{equation}
The {\em individual average-power constraint} is that
\begin{equation}\label{eq:indivaverage}
        \E{     |Z_k^{(t)}|^2 } \le \frac{1}{\beta}, \quad t \in \{ 1,\ldots, \nt \}, \ k\in\Integers.
\end{equation}
The capacity of the channel described in \eqref{eq:MIMOmodel} (or \eqref{eq:MIMOmodel1}) under the individual power constraints \eqref{eq:indivpeak} and \eqref{eq:indivaverage} is denoted as $C_{\text{I}}(\rho, \beta)$ and is given by 
\begin{equation}
        C_{\text{I}}(\rho,\beta) = \lim_{n\to\infty} \frac{1}{n} \sup I(\vect{Z}_1^n; \vect{Y}_1^n)
\end{equation}
with the supremum taken over all distributions on $\vect{Z}_1^n$ satisfying \eqref{eq:indivpeak} and \eqref{eq:indivaverage}. We define
\begin{equation}\label{eq:c_i_beta}
        c_{\text{I}}(\beta) \triangleq \lim_{\rho\downarrow 0} \frac{C_{\text{I}}(\rho,\beta)}{\rho^2}
\end{equation}
when the limit exists.

\subsubsection{Results on MIMO with individual power constraints} 
We have failed to derive the asymptotic capacity of a general MIMO fading channel with individual power constraints. Upper and lower bounds on the asymptotic capacity are presented in Section \ref{sec:MIMOindiv}. Here we present the asymptotic capacity for transmit separable channels.

\begin{proposition}[Asymptotic Capacity---Transmit Separable]\label{prop:MIMOIcapacity}
        If the MIMO channel is transmit separable, then the limit in \eqref{eq:c_i_beta} exists and is given by
        \begin{eqnarray}
                c_{\text{I}}(\beta)  = \frac{1}{2} \left(\sum_{t=1}^{\nt} \alpha_t\right)^2 \max_{0\le a \le \frac{1}{\beta}} \sum_{r=1}^{\nr} \bigg\{ a \lambda_r~~~~~ \nonumber \\ - a^2 R_r^2(0) \bigg\}.\label{eq:MIMOIcapacity}
        \end{eqnarray}
\end{proposition}

The next corollary is a simpler case of the above proposition.

\begin{corollary}[Asymptotic Capacity---Transmit Separable, Nonephemeral, No Average Power Constraint]\label{coro:MIMOInonephemeral}
        If the channel is transmit separable and nonephemeral, and if no average-power constraint is imposed ($\beta=1$), then
        \begin{equation}\label{eq:MIMOInonephemeral}
                c_{\text{I}}(1) = \left( \sum_{t=1}^{\nt} \alpha_t \right)^2 \sum_{r=1}^{\nr} \frac{\lambda_r - R_r^2(0)}{2}.
        \end{equation}
\end{corollary}

\subsubsection{Discussion} 
\begin{itemize}
\item \emph{Input Distributions:} As the proof of Proposition \ref{prop:MIMOIcapacity} shows, an input law that achieves the capacity asymptotically on a transmit separable MIMO channel is to send the same signal on all antennas, with the signal (on each antenna) having the distribution that achieves the low SNR capacity of SISO channels.   If the signal common to the antennas is an FSK signal,  this is the STORM input distribution \cite{SrinivasanVaranasi06}. 
As mention for the SISO channel, the blocklength of the input $n$ is taken to be the blocklength of the channel model in  \cite{SrinivasanVaranasi06}, whereas here we let $n\rightarrow \infty$ to
achieve the maximum capacity asymptote.

\item \emph{Comparison with SISO Channels:} We compare the asymptotic capacity of MIMO channels with individual power constraints to SISO channels. Consider the case when the MIMO channel satisfies (\ref{eq:compareMIMOSISO}) for every antenna pair where $R(\cdot)$ is the autocorrelation function of the SISO channel. Substituting in (\ref{eq:MIMOIcapSeparable}), we get the following expression for the asymptotic capacity of this MIMO channel. 
        \begin{eqnarray}
                c_{\text{I}}(\beta) &=& \frac{\nt^2\nr}{2} \max_{0\le a \le \frac{1}{\beta}} \left\{ a \lambda - a^2 R^2(0)\right\} \\
                &=& \nt^2 \nr \cdot c(\beta).
        \end{eqnarray}
The channel capacity grows linearly with the number of receive antennas ($\nr$) - this is like the case of sum peak power constraints and for similar reasons. The channel capacity grows quadratically with the number of transmit antennas ($\nt$). Increasing the number of transmit antennas reduces the pressure from the peak power constraint because the peak constraint is applied on individual antennas. This causes a gain in capacity. The quadratic dependence stems from the fact that, at vanishingly low peak and average power constraints, the capacity grows quadratically with the power constraints.  A similar observation is made in  \cite{WuSrikant07} for the case that the peak constraint is held fixed as the SNR goes to zero, and a noncoherent block-fading MIMO channel.

\item \emph{Sum and Individual Power Constraints:} We compare the asymptotic capacities of a transmit separable MIMO channel under sum and individual power constraints. The former is given by (Corollary \ref{coro:MIMOSseparable})
        \begin{equation*}
                c_{\text{S}}(\beta) = \frac{\alpha_{\max}^2}{2} \max_{0\le a \le \frac{1}{\beta}} \sum_{r=1}^{\nr} \left\{ a \lambda_r - a^2 R_r^2(0)\right\}.
        \end{equation*}
For the case under individual power constraints, we note that the actually allowed (peak or average) transmit power is $\nt$ times that in the case under sum power constraints. Therefore, we are interested in the value (Proposition \ref{prop:MIMOIcapacity})
        \begin{equation} \label{eq:MIMOIcapSeparable}
                \frac{c_{\text{I}}(\beta)}{\nt^2} = \frac{\alpha_{\text{ave}}^2}{2} \max_{0\le a \le \frac{1}{\beta}} \sum_{r=1}^{\nr} \left\{ a \lambda_r - a^2 R_r^2(0)\right\},
        \end{equation}
where $\alpha_{\text{ave}}$ is the average of $(\alpha_1, \ldots, \alpha_{\nt})$. Thus, the asymptotic capacity under sum power constraints is $(\alpha_{\max}/\alpha_{\text{ave}})^2$ times that under individual power constraints, and is generally larger than the latter. The two values are equal only when all the transmit antennas are equally strong, i.e., $\alpha_1 = \cdots = \alpha_{\nt}$.

\end{itemize}


\subsection{SISO channels with delay spread}\label{sub:delay}

\subsubsection{The Model} 
A SISO channel with delay spread is described as follows. Its time-$k$ complex-valued channel output $Y_k \in \Complex$ is given by
\begin{equation}\label{eq:delaymodel}
        Y_k = \sqrt{\rho} \sum_{t=0}^{T-1} H_k^{(t)} z_{k-t} + W_k,
\end{equation}
where $\sqrt{\rho} z_k$ is the time-$k$ complex-valued channel input; $\{W_k\}$ models the additive noise; and $\{H_k^{(t)}\}$ models the fading in tap $t$. We again assume that $\{W_k\}$ is a sequence of IID random variables of law $\NormalC{0}{1}$. The fading processes are assumed to be independent across the $T$ taps, but correlated in time within each tap, so that the $T$ processes $\{H_k^{(0)}\},\ldots,\{H_k^{(T-1)}\}$ are independent. The autocorrelation function of the fading in Tap $t$ is denoted by $R_t(\cdot)$ and it is assumed that it is square-summable and that it possesses a spectral density function $S_t(\cdot)$. We define
\begin{eqnarray}\label{eq:def_lambda_t}
	\lambda_t & \triangleq & \sum_{\nu=-\infty}^{\infty} |R_t(\nu)|^2 \\
	& = & \int_{-\pi}^{\pi} |S_t(\omega)|^2\frac{\d \omega}{2\pi}. \nonumber
\end{eqnarray}

The following two definitions are analogous to those for MIMO channels.

\begin{definition}
A SISO channel with delay spread is said to be {\em nonephemeral} if the $T$ 
fading processes $\{H_k^{(0)}\},\ldots,\{H_k^{(T-1)}\}$ are all nonephemeral, i.e., if
\begin{equation*}
	\lambda_t \ge 2 R_t^2(0), \quad \text{for all } t \in \{0,\ldots,T-1\}.
\end{equation*}
\end{definition}

\begin{definition}
A SISO channel with delay spread is said to be {\em delay separable} if there are nonnegative constants $\alpha_0, \ldots, \alpha_{T-1}$ and an autocorrelation function $R(\cdot)$ with corresponding spectral density function $S(\cdot)$ such that 
\begin{equation*}
	R_t (k)= \alpha_t R(k)
\end{equation*} 
for all $t \in \{ 0,\ldots, T-1\}$ and $k \in \Integers$. 
\end{definition}

The definition says that a SISO channel with delay spread is delay separable if the fading in all the taps have the same law up to some scaling constants. Note that if a channel is delay separable, then it is nonephemeral if, and only if, $\lambda\ge 2R^2(0)$.

We assume that the input signals are subject to the same constraints as considered earlier for SISO channels with flat fading, i.e., that constraints \eqref{eq:SISOpeak} and \eqref{eq:SISOaverage} are imposed.
The capacity of this channel is denoted as $C_{\text{DS}}(\rho, \beta)$ and is given by
\begin{equation}
        C_{\text{DS}}(\rho,\beta) = \lim_{n\to\infty} \frac{1}{n} \sup I(Z_1^n; Y_1^n),
\end{equation}
where the supremum is taken over all distributions on $Z_1^n$ that satisfy \eqref{eq:SISOpeak} and \eqref{eq:SISOaverage}. We define
\begin{equation}\label{eq:c_ds_beta}
        c_{\text{DS}}(\beta) \triangleq \lim_{\rho\downarrow 0} \frac{C_{\text{DS}}(\rho,\beta)}{\rho^2}
\end{equation}
when the limit exists.

\subsubsection{Results on SISO with Delay Spread}
We identify the asymptotic capacity of SISO channels with delay spread that are delay separable.
\begin{proposition}[Asymptotic Capacity---Delay Separable]\label{prop:delaycapacity}
        If the SISO channel with delay spread \eqref{eq:delaymodel} is delay separable, then the limit in \eqref{eq:c_ds_beta} exists and is given by
        \begin{equation}\label{eq:delaycapacity}
                c_{\text{DS}}(\beta) = \frac{1}{2} \left(\sum_{t=0}^{T-1} \alpha_t\right)^2 \max_{0\le a\le \frac{1}{\beta}} \left\{ a \lambda - a^2 R^2(0) \right\},
        \end{equation}
        where
        \begin{align}
                \lambda & \triangleq \sum_{\nu=-\infty}^\infty |R(\nu)|^2 \\
                & = \int_{-\pi}^\pi |S(\omega)|^2 \frac{\d \omega}{2\pi} \nonumber.
        \end{align}
\end{proposition}

\begin{corollary}[Asymptotic Capacity---Delay Separable, Nonephemeral, No Average Power Constraint]\label{coro:delaynonephemeral}
        If the channel is delay separable and nonephemeral, and if no average-power constraint is imposed ($\beta=1$), then
        \begin{equation}
                c_{\text{DS}}(1) = \left(\sum_{t=0}^{T-1} \alpha_t \right)^2 \frac{\lambda - R^2(0)}{2}.
        \end{equation}
\end{corollary}

\subsubsection{Discussion}
\begin{itemize}
\item \emph{Input Distributions:} As the proof of Proposition \ref{prop:delaycapacity} shows, a signaling scheme that achieves the capacity asymptotically is to send FSK signals with a certain probability, and to send the all-zero signal otherwise. Here the FSK signals can be described as follows. The time-$k$ input is 
\begin{equation*}
        Z_k = \exp ( i \cdot k \Theta)
\end{equation*}
where $i=\sqrt{-1}$, $\Theta$ is a random variable, uniformly distributed over the set $\left\{ \frac{j \cdot 2\pi}{m} : j \in \{0,\ldots,m-1\}\right\}$ for some integer $m>1$. Note that in contrast to the SISO flat fading channels, for SISO channels with delay spread it is in general not optimal to replace FSK with PSK.

\item \emph{Relation with MIMO channels:} An upper bound on $C_{\text{DS}}(\rho, \beta)$ is the capacity of the following multiple-input single-output (MISO) channel with individual power constraints:
\begin{equation}\label{eq:delayMISO}
        Y_k = \sqrt{\rho} \sum_{t=0}^{T-1} H_k^{(t)} z_k^{(t)} + W_k.
\end{equation}
Here $\{W_k\}$ and $\{H_k^{(t)}\}$ are the same as in the SISO channel with delay spread we are considering, and the input signals satisfy $|Z_k^{(t)}|\le 1$ with probability one and $\E{|Z_k^{(t)}|^2} \le \frac{1}{\beta}$ for all $t$. Indeed, with the following additional conditions, the channel \eqref{eq:delayMISO} is the same as \eqref{eq:delaymodel}: 
\begin{equation}\label{eq:MISOadditional}
        z_k^{(t)} = z_{k'}^{(t')} \quad  \text{whenever } k-t = k'-t'.
\end{equation}
Generally speaking, condition \eqref{eq:MISOadditional} is very strong on MISO channels. Therefore, it is usually not optimal to upper bound $C_{\text{DS}}(\rho,\beta)$ by the capacity of the MISO channel with individual power constraints. However, as the proof of Proposition \ref{prop:delaycapacity} shows, this upper bound is tight in the low SNR limit for delay separable channels.
        
\item \emph{Delay Spread does not Waste Energy at Low SNR:} In a delay separable channel, the actually received peak (average) signal power is $\left(\sum_{t=0}^{T-1} \alpha_t\right)$ times the received peak (average) signal power in the corresponding SISO flat fading channel. Proposition \ref{prop:delaycapacity} tells us that at low SNR, the asymptotic capacity of the delay separable channel is the same as that of a SISO flat fading channel with the same received power. Thus, having the power distributed in different taps does not reduce the channel capacity at low SNR. The delay spread channel is similar to a Gaussian channel with noise power which depends on the weighted sum of the past channel input powers. An analogous result for this {\em heating up channel} was observed in \cite{KochLapidoth07}.

\end{itemize}


\section{SISO channels} \label{sec:SISO}

In this section we shall prove the results given in \ref{sub:SISO}. We start with the upper bound. 

\emph{Proof of Proposition \ref{prop:SISOupper}:} To prove that $C(\rho,\beta)\le U(\rho,\beta)$, where $U(\rho,\beta)$ is defined in \eqref{eq:def_U}, it suffices to show that
\begin{equation}\label{eq:SISOproof1}
	\frac{1}{n} I(Z_1^n;Y_1^n) \le U(\rho, \beta)
\end{equation}
for all $n\in\Naturals$ and all distributions on $Z_1^n$ satisfying the peak- and average-power constraints \eqref{eq:SISOpeak} and \eqref{eq:SISOaverage}. To this end, we use the chain rule of mutual information to write
\begin{eqnarray}
	I(Z_1^n; Y_1^n) & = & \sum_{k=1}^n I\left(Z_1^n; Y_k |Y_1^{k-1} \right) \nonumber\\
	& = & \sum_{k=1}^n \left\{ I(Z_1^n, Y_1^{k-1}; Y_k) - I(Y_k; Y_1^{k-1}) \right\} \nonumber\\
	& \le & \sum_{k=1}^n I(Z_1^n, Y_1^{k-1}; Y_k) \nonumber\\
	& = & \sum_{k=1}^n I(Z_1^k, Y_1^{k-1}; Y_k) \label{eq:chain_rule}
\end{eqnarray}
where the last equality follows because the channel has no feedback. To prove \eqref{eq:SISOproof1}, it thus suffices to show that 
\begin{equation}\label{eq:SISOproof2}
	I(Z_1^k, Y_1^{k-1}; Y_k) \le U(\rho, \beta), \quad k\in \Integers.
\end{equation}
By shifting the indices by $-k$ and adding random variables we have
\begin{equation}\label{eq:SISOproof68}
	I(Z_1^k, Y_1^{k-1}; Y_k) \le I(Z_{-\infty}^{0}, Y_{-\infty}^{-1}; Y_0 ).
\end{equation}
It thus follows by \eqref{eq:SISOproof68} that to prove \eqref{eq:SISOproof2} (and hence \eqref{eq:SISOproof1}), it suffices to prove
\begin{equation}\label{eq:SISOproof69}
	I(Z_{-\infty}^{0}, Y_{-\infty}^{-1}; Y_0 ) \le U(\rho,\beta)
\end{equation}
for all distributions on $Z_{-\infty}^0$ satisfying the constraints \eqref{eq:SISOpeak} and \eqref{eq:SISOaverage}. To that end, we write
\begin{equation}\label{eq:SISOproof99}
	I(Z_{-\infty}^{0}, Y_{-\infty}^{-1}; Y_0 ) = h(Y_0) - h\left( Y_0\big| Z_{-\infty}^0, Y_{-\infty}^{-1} \right)
\end{equation}
and bound the two terms on the RHS separately. As to the first term we note that the variance of $Y_0$ is given by
\begin{eqnarray*}
	\E{|Y_0|^2} & = & \E{ \left| \sqrt{\rho} H_0 Z_0 + W_0 \right|^2 } \\
	& = & \E{|W_0|^2} + \E{ \left| \sqrt{\rho} H_0 Z_0 \right|^2 } \\
	& = & 1 + \rho \E{|Z_0|^2},
\end{eqnarray*}
so that the differential entropy of $Y_0$ is bounded by
\begin{equation}\label{eq:SISOproof3}
	h(Y_0) \le \log \pi e \left( 1 + \rho \E{ |Z_0|^2 } \right).
\end{equation}
To study the term $h(Y_0|Z_{-\infty}^0, Y_{-\infty}^{-1})$, we note that when $z_0$ is known, the past channel inputs and outputs provide information about $Y_0$ only through the prediction of $H_0$. So conditional on $z_{-\infty}^0$ and $y_{-\infty}^{-1}$, $Y_0$ has the form
\begin{equation}\label{eq:SISOproof16}
	Y_0 = \sqrt{\rho} \left(\hat{h}_0 + \tilde{H}_0 \right) z_0 + W_0,
\end{equation}
where
\begin{equation}
	\hat{h}_0 = \E{ H_0 | z_{-\infty}^{-1}, y_{-\infty}^{-1} }
\end{equation}
is the conditional expectation of $H_0$ conditional on $\left(z_{-\infty}^{-1}, y_{-\infty}^{-1}\right)$, and where
\begin{equation}
	\tilde{H}_0 = H_0 - \hat{h}_0
\end{equation}
is the error in predicting $H_0$ based on $\left( z_{-\infty}^{-1}, y_{-\infty}^{-1} \right)$. Conditional on $z_{-\infty}^{-1}$, since $Y_{-\infty}^{-1}$ and $H_0$ are jointly PCN, we have that $\tilde{H}_0$ is zero-mean PCN with variance that does not depend on $Y_{-\infty}^{-1}$. We thus conclude that conditional on $\left( z_{-\infty}^0, y_{-\infty}^{-1}\right)$, $Y_0$ is PCN with mean $\sqrt{\rho}\hat{h}_0 z_0$ and variance $\left( \rho \E{ |\tilde{H}_0 |^2 \big| z_{-\infty}^{-1} } |z_0|^2 + 1\right)$. We next show that for all $z_{-\infty}^{-1}$,
\begin{equation}\label{eq:SISOproof4}
	\E{ | \tilde{H}_0 |^2 \big| z_{-\infty}^{-1} } \ge \sigma^2(\rho),
\end{equation}
and therefore
\begin{equation}\label{eq:SISOproof5}
	h(Y_0 | Z_{-\infty}^0, Y_{-\infty}^{-1} ) \ge \E{ \log \pi e \left( 1 + \rho |Z_0|^2 \sigma^2(\rho) \right) }.
\end{equation} 
Inequality \eqref{eq:SISOproof4} is justified by noting that the prediction error (i.e., the variance of $\tilde{H}_0$) is minimized when all the past inputs have maximum amplitude, i.e., $|z_{-1}| = |z_{-2}|\cdots = 1$. In this case, the estimation of $H_0$ based on $\left( Z_{-\infty}^0, Y_{-\infty}^0 \right)$ reduces to estimation based on $\left\{\ldots, H_{-2}+N_{-2}, H_{-1}+N_{-1} \right\}$ where $N_k=\frac{W_k}{Z_k}$, so $\{ N_k \}$ are IID random variables of law $\NormalC{0}{1/\rho}$ \cite{Lapidoth05}. The error of the latter estimation is given by $\sigma^2(\rho)$. Combining \eqref{eq:SISOproof3} and \eqref{eq:SISOproof5} we obtain
\begin{eqnarray}
	I(Z_{-\infty}^0, Y_{-\infty}^{-1}; Y_0 ) \le \log(1+ \rho \E{ |Z_0|^2 } )~~~~~~~~~~~~\nonumber\\
	{} - \E{ \log\left(1+\rho\sigma^2(\rho) |Z_0|^2\right)}. \label{eq:SISOproof6}
\end{eqnarray}
We next continue with the proof of \eqref{eq:SISOproof69} by further upper bounding the RHS of \eqref{eq:SISOproof6}. Let $a \triangleq \E{|Z_0|^2}$. By the average-power constraint \eqref{eq:SISOaverage},
\begin{equation}\label{eq:constraint_a}
	0\le a \le \frac{1}{\beta}. 
\end{equation}
By the concavity of the $\log$ function, the RHS of \eqref{eq:SISOproof6} is maximized over all distributions satisfying the peak constraint \eqref{eq:SISOpeak} and the constraint $\E{|Z_0|^2} = a$ by
\begin{equation*}
	|Z_0|^2 = \begin{cases} 1 & \text{with probability} \ a \\ 0 & \text{with probability} \ 1-a \end{cases}.
\end{equation*}
Consequently, for some $a\in\left[0,\frac{1}{\beta} \right]$,
\begin{eqnarray}
	I(Z_{-\infty}^0, Y_{-\infty}^{-1}; Y_0) & \le & \log(1+\rho a) - a \log(1+\rho \sigma^2(\rho))\nonumber\\
	& = & \log(1+\rho a) - a I (\rho). \label{eq:SISO_proof_jing}
\end{eqnarray}
Maximizing the RHS of \eqref{eq:SISO_proof_jing} over all $a \in \left[ 0, \frac{1}{\beta} \right]$ yields the optimal choice $a^* = \zeta(\rho, \beta)$ and the maximum value $U(\rho, \beta)$, and thus establishes \eqref{eq:SISOproof69}.
\hfill
\IEEEQED
	
\emph{Proof of Proposition \ref{prop:SISOcapacity}:} The proof consists of two parts. The first part shows that 
\begin{equation}\label{eq:SISOproof36}
	\lim_{\rho\downarrow 0}\frac{U(\rho, \beta)}{\rho^2} = \begin{cases}  
    \frac{\lambda^2}{8} &
    \text{if $\displaystyle{\lambda < \frac{2}{\beta}}$} \\
    \frac{\lambda}{2\beta} - \frac{1}{2\beta^2} & \text{if
      $\displaystyle{\lambda \ge \frac{2}{\beta}}$}
   \end{cases}.
\end{equation}
This combines with Proposition \ref{prop:SISOupper} to prove that $c(\beta)$ is upper bounded by the RHS of \eqref{eq:SISOcapacity2}. The second part demonstrates that $c(\beta)$ is also lower bounded by the RHS of \eqref{eq:SISOcapacity2}.

We begin with the first part. By \eqref{eq:IrhoSISO} we have that
\begin{equation}\label{eq:SISOproof77}
	\lim_{\rho\downarrow 0} \left\{\frac{1}{I(\rho)}- \frac{1}{\rho}\right\} = \frac{\lambda}{2}.
\end{equation}
We study two different cases corresponding to $\lambda<\frac{2}{\beta}$ and to $\lambda\ge \frac{2}{\beta}$. For the case $\lambda<\frac{2}{\beta}$, we have by \eqref{eq:def_zeta} and \eqref{eq:SISOproof77}
\begin{equation*}
	\lim_{\rho\downarrow 0} \zeta(\rho, \beta) = \frac{\lambda}{2}. 
\end{equation*}
Thus, in this case,
\begin{eqnarray}
	\lim_{\rho\downarrow 0} \frac{U(\rho,\beta)}{\rho^2} & = & \lim_{\rho\downarrow 0} \frac{\log\left(1+\frac{\lambda}{2}\rho\right) - \frac{\lambda}{2}I(\rho)}{\rho^2} \nonumber\\
	& = & \lim_{\rho\downarrow 0} \frac{1}{\rho^2}\left\{\left(\frac{\lambda}{2}\rho - \frac{\lambda^2}{8} \rho^2 + o(\rho^2) \right) \right. \nonumber\\
	&& ~~~~~~~~~~~~\left. {}- \frac{\lambda}{2}\left(\rho - \frac{\lambda\rho^2}{2} + o(\rho^2) \right)\right\} \nonumber\\
	& = & \frac{\lambda^2}{8},\qquad\qquad \left(\lambda<\frac{2}{\beta}\right),\label{eq:SISOproof78}
\end{eqnarray}
where the second equality follows by a second order Taylor expansion of the $\log$ function and \eqref{eq:IrhoSISO}. In the case where $ \lambda \ge \frac{2}{\beta}$, we have by \eqref{eq:def_zeta} and \eqref{eq:SISOproof77} that
\begin{equation*}
	\lim_{\rho\downarrow 0} \zeta(\rho, \beta) = \frac{1}{\beta},
\end{equation*}
so that
\begin{eqnarray}
	\lim_{\rho\downarrow 0} \frac{U(\rho,\beta)}{\rho^2} & = & \lim_{\rho\downarrow 0} \frac{\log\left(1+\frac{1}{\beta}\rho \right) - \frac{1}{\beta}I(\rho)}{\rho^2} \nonumber\\
	& = & \lim_{\rho\downarrow 0} \frac{1}{\rho^2}\left\{ \left(\frac{1}{\beta}\rho - \frac{1}{2\beta^2}\rho^2 + o(\rho^2) \right) \right. \nonumber\\
	&& ~~~~~~~~~~~~\left. {} - \frac{1}{\beta}\left(\rho - \frac{\lambda \rho^2}{2} + o(\rho^2) \right)\right\}\nonumber\\
	& = & \frac{\lambda}{2\beta} - \frac{1}{2\beta^2},\qquad\qquad \left(\lambda>\frac{2}{\beta} \right). \label{eq:SISOproof79}
\end{eqnarray}
The limits \eqref{eq:SISOproof78} and \eqref{eq:SISOproof79} establish \eqref{eq:SISOproof36}.

We next turn to the second part. We shall now choose a joint distribution on $Z_1^n$ for every $n\in\Naturals$ and show that under this distribution
\begin{equation}\label{eq:SISOproof100}
	\lim_{n\to\infty} \lim_{\rho\downarrow 0} \frac{\frac{1}{n}I(Z_1^n; Y_1^n)}{\rho^2} = \frac{1}{2} \max_{0\le a \le \frac{1}{\beta} } \left\{ a\lambda - a^2\right\},
\end{equation}
where the RHS of \eqref{eq:SISOproof100} is equal to the RHS of \eqref{eq:SISOcapacity} (or \eqref{eq:SISOcapacity2}). The  expression on the LHS of \eqref{eq:SISOproof100} indeed forms a lower bound on $c(\beta)$ because, by Lemma \ref{lem:lower}, for any $n\in\Naturals$ and any distribution on $Z_1^n$ satisfying the peak- and average-power constraints \eqref{eq:SISOpeak} and \eqref{eq:SISOaverage}, 
\begin{equation}
	\frac{1}{n}I(Z_1^n;Y_1^n)\le C(\rho,\beta),
\end{equation}
and therefore, for any $n \in \Integers$
\begin{equation*}
	\liminf_{\rho\downarrow 0} \frac{\frac{1}{n}I(Z_1^n; Y_1^n)}{\rho^2} \le \liminf_{\rho\downarrow 0} \frac{C(\rho,\beta)}{\rho^2}.
\end{equation*}
This inequality also holds in the limit as $n \to \infty$. 

For a fixed $n$, the proposed distribution on $Z_1^n$ can be described as follows:
\begin{equation*}
	Z_k = U \cdot \Phi_k, \quad k \in \{1,\ldots, n\},
\end{equation*}
where
\begin{equation*}
	U=\begin{cases} 1 & \text{with probability } a \\ 0 & \text{with probability } 1-a \end{cases}
\end{equation*}
for some $0\le a \le\frac{1}{\beta}$, and $\Phi_1^n$ are random variables satisfying
\begin{equation*}
	|\Phi_k| = 1,\ \E{\Phi_k} = 0, \quad k \in \{1,\ldots, n\}
\end{equation*}
and
\begin{equation*}
	\E{ \Phi_k \Phi_{k'}^* } = 0, \quad k\neq k'.
\end{equation*}
Examples of distributions on $\{ \Phi_k \}$ have been given in Section \ref{sec:mainresults}. Under the proposed distribution on $Z_1^n$, the mutual information $I (Z_1^n ; Y_1^n)$ when $\rho$ is small can be computed by applying \cite[Corollary 1]{PrelovVerdu04}, which is restated in this paper as Lemma \ref{lemm:PrelovVerdu}. We apply the lemma by letting $\vect{Z} = \trans{(Z_1,\ldots, Z_n)}$ and $\vmat{H}$ be the diagonal matrix with diagonal entries $H_1,\ldots,H_n$. The calculation shows that
\begin{equation}\label{eq:SISOproof844}
	\frac{1}{n}I(Z_1^n; Y_1^n) =  \frac{\rho^2}{2} \left( a\cdot \frac{1}{n} \sum_{1\le i, j \le n} |R(i-j)|^2 - a^2  \right) + o(\rho^2).
\end{equation}
Noting that by \eqref{eq:def_lambda}
\begin{eqnarray}
	\lim_{n\to\infty} \frac{1}{n} \sum_{1\le i, j \le n} |R(i-j)|^2 & = & \lim_{n\to\infty} \frac{1}{n} \sum_{\nu=-(n-1)}^{n-1} \left(n-|\nu|\right) \left| R(\nu) \right|^2 \nonumber\\
	& = & \lim_{n\to \infty} \sum_{\nu=-(n-1)}^{n-1} \left(1-\frac{|\nu|}{n} \right) \left|R(\nu)\right|^2 \nonumber\\
	& = &\lambda, \label{eq:SISO_proof_87}.
\end{eqnarray}
we obtain from \eqref{eq:SISOproof844} that
\begin{equation}\label{eq:SISOproof17}
	\lim_{n\to\infty} \lim_{\rho\downarrow 0} \frac{\frac{1}{n} I(Z_1^n; Y_1^n)}{\rho^2} = \frac{1}{2}\left(\lambda a - a^2\right).
\end{equation}
Equality \eqref{eq:SISOproof100} follows when we choose $a^*\in \left[0,\frac{1}{\beta}\right]$ that maximizes the RHS of \eqref{eq:SISOproof17}.
\hfill
\IEEEQED

We shall now prove Proposition \ref{prop:SISOiid}. Before doing so, we present a lemma which studies the problem of predicting the current fade $H_0$ based on the past channel inputs and outputs. We have seen in the proof of Proposition \ref{prop:SISOupper} that if all the past inputs satisfy $|z_k|=1$, $k \in \{ \ldots,-2,-1\}$, then this problem is reduced to predicting $H_0$ based on a noisy observation of the past $H_k + N_k$, $k \in \{ \ldots,-2,-1\}$, where $\{N_k\}$ is a sequence of IID PCN noise. As shown in the next lemma, this problem becomes more difficult when $|z_k|$ is not always 1.

\begin{lemma}\label{lemm:predictZY}
	If the autocorrelation function $R(\cdot)$ of the unit-variate fading process $\{H_k\}$ is absolutely summable, and if the input symbols $\{ Z_k \}$ satisfy the peak-power constraint \eqref{eq:SISOpeak}, then the conditional distribution of $H_0$ conditional on the past inputs $Z_{-\infty}^{-1}= z_{-\infty}^{-1}$ and outputs $Y_{-\infty}^{-1}= y_{-\infty}^{-1}$ is PCN with a variance $\varsigma^2(\rho, z_{-\infty}^{-1})$ which does not depend on $y_{-\infty}^{-1}$, and $\varsigma^2(\rho, z_{-\infty}^{-1})$ satisfies
	\begin{equation}\label{eq:predictZY}
		\varsigma^2(\rho, z_{-\infty}^{-1}) = 1 - \rho \sum_{\nu=-\infty}^{-1} |R(\nu)|^2 |z_\nu|^2 + o(\rho),
	\end{equation}
	where $o(\rho)$ is uniform in $z_{-\infty}^{-1}$.
\end{lemma}
\begin{proof}
See Appendix \ref{app:predictZY}.
\end{proof}

\emph{Proof of Proposition \ref{prop:SISOiid}:}
	Since we are interested in the fact that IID inputs do not generally achieve the channel capacity at low SNR, we shall concentrate on the proof of the upper bound. The achievability part can be proved by choosing an IID distribution on $Z_1^n$ taking values in $\{0,\pm1\}$ and by then applying Lemma \ref{lemm:PrelovVerdu}. 
	
	As in the proof of Proposition \ref{prop:SISOupper}, to show that
	\begin{equation}\label{eq:SISOproof83}
		c_{\text{IID}}(\beta) \le \begin{cases} \frac{1}{8(2-\lambda)} & \text{if } \lambda<2-\frac{\beta}{2} \\ \frac{1}{2\beta}+ \frac{\lambda-2}{2\beta^2} & \text{if } \lambda \ge 2-\frac{\beta}{2} \end{cases},
	\end{equation}
	it suffices to show that for all IID distributions on $Z_{-\infty}^{0}$ satisfying the peak- and average-power constraints \eqref{eq:SISOpeak} and \eqref{eq:SISOaverage}, 
	\begin{equation}\label{eq:SISOproof84}
		\limsup_{\rho\downarrow 0} \frac{I(Z_{-\infty}^{0}, Y_{-\infty}^{-1}; Y_0)}{\rho^2} \le \begin{cases} \frac{1}{8(2-\lambda)} & \text{if } \lambda<2-\frac{\beta}{2} \\ \frac{1}{2\beta}+ \frac{\lambda-2}{2\beta^2} & \text{if } \lambda \ge 2-\frac{\beta}{2} \end{cases}.
	\end{equation}
	To prove \eqref{eq:SISOproof84}, we decompose $I(Z_{-\infty}^{0}, Y_{-\infty}^{-1}; Y_0)$ as in \eqref{eq:SISOproof99} and treat the two terms on the RHS separately. The first term satisfies
	\begin{equation}\label{eq:SISOproof91}
		h(Y_0) \le \log \pi e \left( 1 + \rho \E{ |Z_0|^2 } \right).
	\end{equation}
	To study the second term, we note that given $Z_0$, the past channel inputs and outputs provide information about $Y_0$ only through the prediction of $H_0$. Denoting
	\begin{equation*}
		\hat{h}_0 \triangleq \E{H_0 | z_{-\infty}^{-1}, y_{-\infty}^{-1} },
	\end{equation*}
	we can express $Y_0$ in the same form as \eqref{eq:SISOproof16}.
	By Lemma \ref{lemm:predictZY}, given $\hat{h}_0$ and $z_0$, the distribution of $H_0$ is PCN of variance $\varsigma^2(\rho, z_{-\infty}^{-1})$, thus the distribution of $Y_0$ is PCN of variance $\left(1+\rho |z_0|^2 \varsigma^2(\rho, z_{-\infty}^{-1})\right)$. So we have
	\begin{equation}\label{eq:SISOproof92}
		h(Y_0|Z_0, \hat{H}_0) = \E{ \log \pi e \left(1 +  \rho |Z_0|^2 \varsigma^2(\rho, Z_{-\infty}^{-1}) \right) }.
	\end{equation}
	In the following calculations, let $a \triangleq \E{|Z_0|^2}$. Note that since $\{ Z_k\}$ are IID, $a = \E{|Z_k|^2}$ for all $k$. We obtain from \eqref{eq:SISOproof99}, \eqref{eq:SISOproof91} and \eqref{eq:SISOproof92} that
	\begin{eqnarray}
		I(Z_{-\infty}^{0}, Y_{-\infty}^{-1}; Y_0) & \le & \log \pi e \left(1 + \rho \E{|Z_0|^2} \right) - \E{ \log \pi e \left( 1 + \rho |Z_0|^2 \varsigma^2(\rho, Z_{-\infty}^{-1}) \right)} \nonumber\\
		& = & \left( \frac{\lambda-2}{2} a^2 + \frac{1}{2} \E{|Z_0|^4} \right) \rho^2 + o(\rho^2) \nonumber\\
		& \le & \left( \frac{\lambda-2}{2} a^2 + \frac{1}{2} a \right) \rho^2 + o(\rho^2), \label{eq:SISOproof93}
	\end{eqnarray}
	where the equality follows by calculations using the first order Taylor Expansion of $\rho \mapsto \varsigma^2(\rho, z_{-\infty}^{-1})$ (Lemma \ref{lemm:predictZY}), the second order Taylor Expansion of $x \mapsto \log(1+x)$, and the fact that $\{Z_k\}$ are IID; the last inequality follows because when $|Z_0|\le 1$, $\E{|Z_0|^4} \le \E{|Z_0|^2} = a$. 

	From \eqref{eq:SISOproof93} it follows that
	\begin{equation}\label{eq:SISOproof94}
		\limsup_{\rho\downarrow 0} \frac{I(Z_{-\infty}^{0}, Y_{-\infty}^{-1}; Y_0)}{\rho^2} \le \max_{0 \le a \le \frac{1}{\beta}} \left\{ \frac{\lambda - 2}{2} a ^2 + \frac{1}{2} a \right\}.
	\end{equation}
	Inequality \eqref{eq:SISOproof84} (and thus \eqref{eq:SISOproof83}) follows from \eqref{eq:SISOproof94} because when $\lambda \ge 2 - \frac{\beta}{2}$, the maximum of the RHS of \eqref{eq:SISOproof94} is $\frac{1}{2\beta} + \frac{\lambda-2}{2\beta^2}$ and is achieved by $a = \frac{1}{\beta}$; when $\lambda < 2 - \frac{\beta}{2}$, the maximum is $\frac{1}{8(2-\lambda)}$ and is achieved by $a=\frac{1}{2(2-\lambda)}$
\hfill
\IEEEQED


\section{MIMO channels with sum power constraints} \label{sec:MIMOsum}

In this section we shall prove the results on MIMO channels with sum power constraints. We shall first prove the upper bound (Proposition \ref{prop:MIMOSupper}) in two special cases, namely, for MISO channels and for single-input multiple-output (SIMO) channels, and then combine the proofs of these two special cases to prove Proposition \ref{prop:MIMOSupper} generally for MIMO channels. The asymptotic capacity results will then be proved with the help of Proposition \ref{prop:MIMOSupper}.

We start with upper bound on the capacity of the SIMO channel. Consider the channel \eqref{eq:MIMOmodel1} with $\nt = 1$. We drop the superscript $(t)$ and rewrite the channel as
\begin{equation}
	Y_k^{(r)} = \sqrt{\rho} H_k^{(r)} Z_k + W_k^{(r)},\quad r\in\{1,\ldots,\nr\}, \ k \in \Integers.
\end{equation}
Similarly, below we write $I_r(\rho)$ instead of $I_{1,r}(\rho).$
The sum power constraints reduce to constraints on the scalars $\{Z_k\}$:
\begin{eqnarray}
	|Z_k| &\le &1, \label{eq:SIMO_peak}\\
	\E{|Z_k|^2} & \le & \frac{1}{\beta}. \label{eq:SIMO_average}
\end{eqnarray}
We denote the capacity of this channel by $C_{\text{SIMO-S}}(\rho, \beta)$. For this SIMO channel Proposition \ref{prop:MIMOSupper} reduces to
\begin{lemma}\label{lemm:SIMOSupper}
	The capacity $C_{\text{SIMO-S}}(\rho, \beta)$ is upper bounded by
	\begin{equation}\label{eq:SIMOS_upper}
		C_{\text{SIMO-S}}(\rho, \beta) \le U_{\text{SIMO-S}}(\rho,\beta),
	\end{equation}
	where
	\begin{equation}
		U_{\text{SIMO-S}}(\rho,\beta) \triangleq \max_{0\le a\le \frac{1}{\beta}} \sum_{r=1}^{\nr} \left\{ \log(1+ a \rho R_r(0)) - a I_r(\rho) \right\}.
	\end{equation}
\end{lemma}

\begin{proof}
	Analogously to \eqref{eq:SISOproof69}, to prove \eqref{eq:SIMOS_upper} it suffices to show that
	\begin{equation}\label{eq:MIMOS_proof_99}
		I(Z_{-\infty}^0, \vect{Y}_{-\infty}^{-1}; \vect{Y}_0 ) \le U_{\text{SIMO-S}}(\rho,\beta)
	\end{equation}
	for all distributions on $Z_{-\infty}^0$ satisfying \eqref{eq:SIMO_peak} and \eqref{eq:SIMO_average}. To prove \eqref{eq:MIMOS_proof_99}, we expand its LHS as
	\begin{equation}\label{eq:MIMOS_proof_100}
		I(Z_{-\infty}^0, \vect{Y}_{-\infty}^{-1}; \vect{Y}_0 ) = h(\vect{Y}_0) - h(\vect{Y}_0 | Z_{-\infty}^0, \vect{Y}_{-\infty}^{-1} ),
	\end{equation}
	and proceed to bound the two terms separately. For $h(\vect{Y}_0)$ we have
	\begin{equation}\label{eq:MIMOS_proof_101}
		h(\vect{Y}_0) = h(Y_0^{(1)}, \ldots, Y_0^{(\nr)} ) \le \sum_{r=1}^{\nr} h(Y_0^{(r)} ).
	\end{equation}
	We now consider $h(\vect{Y}_0 | Z_{-\infty}^0, \vect{Y}_{-\infty}^{-1} )$. Because there is no dependence between the $\nr$ fading processes $\{H_k^{(r)} \}$, we have that, conditional on $Z_{-\infty}^0$ and $\vect{Y}_{-\infty}^{-1}$, the random variables $Y_0^{(1)},\ldots, Y_0^{(\nr)}$ are mutually independent. Therefore,
	\begin{eqnarray}
		h(\vect{Y}_0 | Z_{-\infty}^0, \vect{Y}_{-\infty}^{-1} )& = &\sum_{r=1}^{\nr} h\left(Y_0^{(r)} | \vect{Y}_{-\infty}^{-1}, Z_{-\infty}^0 \right)\nonumber\\
		& = & \sum_{r=1}^{\nr} h\left(Y_0^{(r)} | (Y^{(r)})_{-\infty}^{-1}, Z_{-\infty}^0 \right)\label{eq:MIMOS_proof_102}
	\end{eqnarray}
	where the second equality follows because, conditional on $(Y^{(r)})_{-\infty}^{-1}$ and $Z_{-\infty}^0$, the signal at the $r$-th receive antenna $Y_0^{(r)}$ is independent of the past outputs of other antennas. From \eqref{eq:MIMOS_proof_100}, \eqref{eq:MIMOS_proof_101} and \eqref{eq:MIMOS_proof_102} we have
	\begin{eqnarray}
		I(Z_{-\infty}^0, \vect{Y}_{-\infty}^{-1}; \vect{Y}_0) & \le & \sum_{r=1}^{\nr}\left\{h(Y_0^{(r)}) - h\left(Y_0^{(r)} | (Y^{(r)})_{-\infty}^{-1}, Z_{-\infty}^0 \right) \right\}\nonumber\\
		& = & \sum_{r=1}^{\nr} I\left(Z_{-\infty}^{0}, \left(Y^{(r)}\right)_{-\infty}^{-1}; Y_0^{(r)} \right).\label{eq:MIMOSproof1}
	\end{eqnarray}
	For every $r\in\left\{1,\ldots,\nr\right\}$, the value of $I\left(Z_{-\infty}^{0}, \left(Y^{(r)}\right)_{-\infty}^{-1}; Y_0^{(r)} \right)$ can be upper-bounded in the same way as \eqref{eq:SISO_proof_jing}, thus we have from \eqref{eq:MIMOSproof1}
	\begin{equation*}
		I(Z_{-\infty}^0, \vect{Y}_{-\infty}^{-1}; \vect{Y}_0) \le \sum_{r=1}^{\nr} \left\{ \log(1+\rho R_r(0)) a - a I_r(\rho) \right\},
	\end{equation*}
	where $a \triangleq \E{|X_0|^2}$. Maximizing the RHS of this inequality over $a$ yields \eqref{eq:MIMOS_proof_99}.
\end{proof}

We now turn to the MISO channel. Consider the channel model in \eqref{eq:MIMOmodel1} when $\nr = 1$. Dropping the superscript $(r)$ we rewrite the channel as
\begin{equation}
	Y_k = \sqrt{\rho} \sum_{t=1}^{\nt} H_k^{(t)} Z_k^{(t)} + W_k.
\end{equation}
Similarly, below we write $I_t(\rho)$ instead of $I_{t,1}(\rho),$ and $\sigma^2_{t}(\rho)$ instead of
$\sigma^2_{t,1}(\rho)$.
Denote the capacity of this channel under the sum constraints \eqref{eq:sumpeak} and \eqref{eq:sumaverage} by $C_{\text{MISO-S}}(\rho,\beta)$. Proposition \ref{prop:MIMOSupper} reduces to the following lemma.
\begin{lemma}\label{lemm:SMISO}
	The capacity $C_{\text{MISO-S}}(\rho,\beta)$ is upper bounded by
	\begin{equation}\label{eq:MISOS_upper}
		C_{\text{MISO-S}}(\rho,\beta) \le U_{\text{MISO-S}}(\rho,\beta)
	\end{equation}
	where
	\begin{eqnarray}
		U_{\text{MISO-S}}(\rho,\beta) \triangleq \max_{\vect{a} \in \set{A}(\beta)} \left\{ \log\left(1+ \rho \sum_{t=1}^{\nt} R_t(0) a_t\right)\right.~~~~~~~~ \nonumber
		\\ \left. - \sum_{k=1}^{\nt} a_t I_t(\rho) \right\}.
	\end{eqnarray}
\end{lemma}
\begin{proof}
	In analogy to the SISO case, to prove \eqref{eq:MISOS_upper} it suffices to show that
	\begin{equation}\label{eq:MISOS_proof_105}
		I(\vect{Z}_{-\infty}^0, Y_{-\infty}^{-1}; Y_0) \le U_{\text{MISO-S}}(\rho,\beta)
	\end{equation}
	for all input distributions satisfying \eqref{eq:sumpeak} and \eqref{eq:sumaverage}. To prove \eqref{eq:MISOS_proof_105}, we expand its LHS as
	\begin{equation}\label{eq:MISOS_proof_106}
		I(\vect{Z}_{-\infty}^0, Y_{-\infty}^{-1}; Y_0) = h(Y_0) - h(Y_0 | \vect{Z}_{-\infty}^0, Y_{-\infty}^{-1})
	\end{equation}
	and bound the two terms on the RHS of \eqref{eq:MISOS_proof_106} separately.
	As in the SISO case, $h(Y_0)$ is upper bounded by the differential entropy of a PCN random variable with the same variance as $Y_0$. The variance of $Y_0$ is given by
	\begin{equation*}
		\E{|Y_0|^2} = \E{|W_k|^2} + \sum_{t=1}^{\nt}\E{|\sqrt{\rho} H_k^{(t)}Z_k^{(t)}|^2} = 1 + \rho \sum_{t=1}^{\nt} R_t(0) \E{|Z_0^{(t)}|^2}.
	\end{equation*}
	Hence,
	\begin{equation}\label{eq:MISOS_proof_107}
		h(Y_0) \le \log \left( \pi e \left( 1+ \rho \sum_{t=1}^{\nt} R_t(0) \E{|Z_0^{(t)}|^2} \right) \right).
	\end{equation}
	We now consider the term $h(Y_0 | \vect{Z}_{-\infty}^0, Y_{-\infty}^{-1})$. To bound its value, we consider for every $k\in\Integers$ the random variable $W_k'$ given by
	\begin{equation}\label{eq:SdecomposeW}
		W_k' = \sum_{t=1}^{\nt} N_{k}^{(t)}  Z_k^{(t)} + \sqrt{ 1- \| \vect{Z}_k \|_2^2 } N_k', 
	\end{equation}
	where $\{N_{k}^{(1)}, \ldots, N_{k}^{(\nt)}\}$ and $N_k'$ are IID random variables of law $\NormalC{0}{1}$, which are independent of the channel inputs and fading processes. It is easy to justify that, with our definition, $\{W_k'\}$ is a sequence of IID $\NormalC{0}{1}$ random variables independent of the channel inputs and the fading processes. Thus we may replace the additive noise $W_k$ with $W_k'$ for every $k\in\Integers$ in the channel model without actually changing the channel law. When we do this, the time-$k$ output $Y_k$ can be written as
	\begin{eqnarray}
		Y_k = \sum_{t=1}^{\nt} (\sqrt{\rho} H_k^{(t)} + N_{k}^{(t)}) Z_k^{(t)} ~~~~~~~~~~~\nonumber\\
		+ \sqrt{1-\sum_{k=1}^{\nt} |Z_k^{(t)}|^2 } N_k'. \label{eq:SdecomposeY}
	\end{eqnarray}
	Conditional on $\vect{Z}_0$ and on $(\sqrt{\rho} H_k^{(t)}+N_{k}^{(t)})$ for all $k\in \{\ldots,-2,-1\}$ and $t\in\{1,\ldots,\nt\}$, the current output $Y_0$ is independent of $\vect{Z}_{-\infty}^{-1}$ and $Y_{-\infty}^{-1}$. Thus we have
	\begin{equation}\label{eq:MISOS_proof_114}
		h(Y_0 | \vect{Z}_{-\infty}^0, Y_{-\infty}^{-1}) \ge h\left(Y_0 | \vect{Z}_0, (\sqrt{\rho}\vect{H}_k + \vect{N}_k)_{k=-\infty}^{-1}\right).
	\end{equation}
	Conditional on $\vect{Z}_0$, for every $t\in\{1,\ldots,\nt\}$, the values of $\left( \sqrt{\rho} H_k^{(t)} + N_k^{(t)} \right)_{k=-\infty}^{-1}$ provide information about $Y_0$ only through the prediction of $H_0^{(t)}$. Furthermore, this prediction (and, in particular, the prediction error) is independent between different $t$'s. The error in predicting $H_0^{(t)}$ based on $\left( \sqrt{\rho} H_k^{(t)} + N_k^{(t)} \right)_{k=-\infty}^{-1}$ is $\sigma_t^2(\rho)$. We thus obtain that, conditional on $\vect{Z}_0$ and $(\sqrt{\rho}\vect{H}_k + \vect{N}_k)_{k=-\infty}^{-1}$, the random variable $Y_0$ is PCN with variance $1 + \rho \sum_{t=1}^{\nt} |Z_0^{(t)}|^2\sigma_t^2(\rho) $. Consequently, the conditional differential entropy $h\left(Y_0 | \vect{Z}_0, (\sqrt{\rho}\vect{H}_k + \vect{N}_k)_{k=-\infty}^{-1}\right)$ is
	\begin{eqnarray}
		h\left(Y_0 | \vect{Z}_0, (\sqrt{\rho}\vect{H}_k + \vect{N}_k)_{k=-\infty}^{-1}\right) = ~~~~~~~~~~~~~~~~~~~~~~~~~~\nonumber\\ \E{ \log \pi e \left(1+\rho \sum_{t=1}^{\nt} |Z_0^{(t)}|^2\sigma_t^2(\rho) \right)}.\label{eq:MISOS_proof_110}
	\end{eqnarray}
	From \eqref{eq:MISOS_proof_106}, \eqref{eq:MISOS_proof_107}, \eqref{eq:MISOS_proof_114} and \eqref{eq:MISOS_proof_110} it follows that
	\begin{eqnarray}
		I(\vect{Z}_{-\infty}^0, Y_{-\infty}^{-1}; Y_0) \le \log \left( 1+ \sum_{t=1}^{\nt} R_t(0) \E{|Z_0^{(t)}|^2} \right)~~ \nonumber \\  {}- \E{ \log \left(1+\rho \sum_{t=1}^{\nt} |Z_0^{(t)}|^2\sigma_t^2(\rho) \right)}. \label{eq:Suppermutual}
	\end{eqnarray}
	We shall now maximize the RHS of the inequality over the distribution on $\vect{Z}_0$. Let $a_t \triangleq \E{|Z_0^{(t)}|^2}$ for all $t \in \left\{ 1,\ldots, \nt \right\}$. Due to the concavity of the $\log$ function, the expectation of the $\log$ on the RHS of \eqref{eq:Suppermutual} is minimized when for all $t \in \{ 1,\ldots, \nt\}$, with probability $a_t$, 
	\begin{eqnarray*}
		Z_0^{(t)} = 1, \\
		Z_0^{(t')}=0, && t'\neq t,
	\end{eqnarray*}
	and with probability $\left(1-\sum_{t=1}^{\nt} a_t\right)$, $\vect{Z}_0 = \vect{0}$. This minimum value of the expectation of the $\log$ is $\sum_{k=1}^{\nt} a_t I_t(\rho)$. Thus we have from \eqref{eq:Suppermutual}
	\begin{equation}\label{eq:MISOS_proof_1155}
		I(\vect{Z}_{-\infty}^0, Y_{-\infty}^{-1}; Y_0) \le \log\left(1+ \rho \sum_{t=1}^{\nt} R_t(0) a_t\right) - \sum_{k=1}^{\nt} a_t I_t(\rho) .
	\end{equation}
	To prove \eqref{eq:MISOS_proof_105}, it remains to maximize the RHS of the above inequality over $\vect{a}$. Note that due to the peak- and average-power constraints \eqref{eq:sumpeak} and \eqref{eq:sumaverage}, $\vect{a} \in \set{A}(\beta)$ where $\set{A}(\beta)$ is defined in \eqref{eq:defAbeta}. Thus the maximum value of the RHS of \eqref{eq:MISOS_proof_1155} is $U_{\text{MISO-S}}(\rho, \beta)$.
\end{proof}

We now turn to prove the upper bound on the capacity of the MIMO channel with sum power constraints.

\emph{Proof of Proposition \ref{prop:MIMOSupper}:} 
	As in the SISO case, to show $C_\text{S}(\rho,\beta)\le U_\text{S}(\rho,\beta)$, it suffices to prove
	\begin{equation}\label{eq:MIMOS_proof_115}
		I(\vect{Z}_{-\infty}^0, \vect{Y}_{-\infty}^{-1}; \vect{Y}_0) \le U_{\text{S}}(\rho,\beta)
	\end{equation}
	for all distributions on $\vect{Z}_{-\infty}^0$ satisfying \eqref{eq:sumpeak} and \eqref{eq:sumaverage}. As in \eqref{eq:MIMOSproof1} for the SIMO channel, the LHS of \eqref{eq:MIMOS_proof_115} is upper bounded by
	\begin{equation}\label{eq:MIMOS_proof_116}
		I(\vect{Z}_{-\infty}^0, \vect{Y}_{-\infty}^{-1}; \vect{Y}_0) \le \sum_{r=1}^{\nr} I(\vect{Z}_{-\infty}^0, (Y^{(r)})_{-\infty}^{-1}; Y_0^{(r)}).
	\end{equation}
	By \eqref{eq:MISOS_proof_1155}, when fixing $\E{|Z_0^{(t)}|^2} = a_t$ for all $t$, each summand on the RHS of \eqref{eq:MIMOS_proof_116} is upper bounded by
	\begin{equation} \label{eq:MIMOS_proof_117}
		I(\vect{Z}_{-\infty}^0, (Y^{(r)})_{-\infty}^{-1}; Y_0^{(r)}) \le \log\left(1+ \rho \sum_{t=1}^{\nt} R_{r,t}(0) a_t\right) - \sum_{k=1}^{\nt} a_t I_{r,t}(\rho).
	\end{equation}
	Thus we obtain from \eqref{eq:MIMOS_proof_116} and \eqref{eq:MIMOS_proof_117} that
	\begin{equation}\label{eq:MIMOS_proof_118}
		I(\vect{Z}_{-\infty}^0, \vect{Y}_{-\infty}^{-1}; \vect{Y}_0) \le \sum_{r=1}^{\nr} \left\{\log\left(1+ \rho \sum_{t=1}^{\nt} R_{r,t}(0) a_t\right) - \sum_{k=1}^{\nt} a_t I_{r,t}(\rho)\right\}.
	\end{equation}
	Note that for input distributions satisfying \eqref{eq:sumpeak} and \eqref{eq:sumaverage} we have $\vect{a}\in\set{A}(\beta)$. Thus maximizing of the RHS of \eqref{eq:MIMOS_proof_118} on $\vect{a}$ yields the value $U_\text{S}(\rho,\beta)$. This establishes \eqref{eq:MIMOS_proof_115}. 
\hfill
\IEEEQED

With the upper bound established, we now proceed to prove the results on the capacity asymptote.

\emph{Proof of Proposition \ref{prop:MIMOScapacity}:}
	The proof consists of two parts. The first part shows that
	\begin{eqnarray}
		\lim_{\rho\downarrow 0} \frac{U_{\text{S}}(\rho,\beta)}{\rho^2}~~~~~~~~~~~~~~~~~~~~~~~~~~~~~~~~~~~~~~~~~~~~~~~~~  \nonumber \\
		 = \frac{1}{2} \max_{\vect{a}\in \set{A}(\beta)} \sum_{r=1}^{\nr} \left\{ \sum_{t=1}^{\nt} a_t \lambda_{r,t} - \left( \sum_{t=1}^{\nt} R_{r,t}(0) a_t \right)^2 \right\}. \label{eq:MIMOS_proof_120}
	\end{eqnarray}
	It then follows from Proposition \ref{prop:MIMOSupper} and \eqref{eq:MIMOS_proof_120} that
	\begin{equation}\label{eq:MIMOS_proof_123}
		\limsup_{\rho\downarrow 0} \frac{C_{\text{S}}(\rho,\beta)}{\rho^2} \le \frac{1}{2} \max_{\vect{a}\in \set{A}(\beta)} \sum_{r=1}^{\nr} \left\{ \sum_{t=1}^{\nt} a_t \lambda_{r,t} - \left( \sum_{t=1}^{\nt} R_{r,t}(0) a_t \right)^2 \right\}.
	\end{equation}
	The second part of the proof shows that the RHS of \eqref{eq:MIMOScapacity} (which is the same as the RHS of \eqref{eq:MIMOS_proof_120}) also forms a lower bound on $\liminf_{\rho\downarrow 0} \frac{C_\text{S}(\rho,\beta)}{\rho^2}$.
	
	To prove \eqref{eq:MIMOS_proof_120}, we use the second order Taylor Expansion of the $\log$ function and the second order Taylor Expansion of the function $I_{r,t}(\cdot)$ \eqref{eq:I_rt_rho} to obtain
	\begin{eqnarray}
		\log\left(1+ \rho \sum_{t=1}^{\nt} R_{r,t}(0) a_t\right) - \sum_{k=1}^{\nt} a_t I_{r,t}(\rho)  ~~~~~~~~~~~~~~~~~~~~~\nonumber \\
		= \frac{\rho^2}{2} \left( \sum_{t=1}^{\nt} a_t \lambda_{r,t} - \left( \sum_{t=1}^{\nt} R_{r,t}(0) a_t \right)^2 \right) + o(\rho^2) \label{eq:MIMOS_proof_121}
	\end{eqnarray}
	where the term $o(\rho^2)$ is uniform in $\vect{a}$. Now \eqref{eq:MIMOS_proof_120} follows by \eqref{eq:MIMOS_proof_121} and \eqref{eq:MIMOSupper}.
	
	We now start the second part of the proof. To derive a lower bound on the asymptotic capacity, we shall find a distribution on $\vect{Z}_1^n$ for every $n\in\Naturals$, such that
	\begin{equation}\label{eq:MIMOS_proof_122}
		\lim_{n \to \infty} \lim_{\rho\downarrow 0} \frac{\frac{1}{n} I(\vect{Z}_1^n; \vect{Y}_1^n)}{\rho^2} = \frac{1}{2} \max_{\vect{a}\in \set{A}(\beta)} \sum_{r=1}^{\nr} \left\{ \sum_{t=1}^{\nt} a_t \lambda_{r,t} - \left( \sum_{t=1}^{\nt} R_{r,t}(0) a_t \right)^2 \right\}.
	\end{equation}
	Note that by Lemma \ref{lem:lower}, the LHS of \eqref{eq:MIMOS_proof_122} forms a lower bound on $\liminf_{\rho\downarrow 0} \frac{C_\text{S}(\rho,\beta)}{\rho^2}$. This combined with \eqref{eq:MIMOS_proof_123} proves \eqref{eq:MIMOS_proof_120}.
	
	For every $n\in\Naturals$ and every vector $\vect{a}\in \set{A}(\beta)$, consider the input distribution
	\begin{equation*}
		Z_k^{(t)} = U^{(t)} \cdot \Phi_k,
	\end{equation*}
	where the random variables $\{U^{(1)},\ldots,U^{(\nt)}\}$ are chosen such that with probability $a_t$,
	\begin{eqnarray*}
		&& U^{(t)} = 1, \\
		&& U^{(t')} = 0, \quad t'\neq t,
	\end{eqnarray*}
	and with probability $(1-\sum_{t=1}^{\nt} a_t)$,
	\begin{equation*}
		\vect{U}=0.
	\end{equation*}
	The random variables $\{ \Phi_k \}$ are chosen in the same way as for the SISO channels, i.e., they satisfy
	\begin{equation*}
		|\Phi_k| = 1, \quad k \in \{1,\ldots,n\}
	\end{equation*}
	and
	\begin{equation*}
		\E{\Phi_k \Phi_{k'}^{*} } =0, \quad k \neq k'.
	\end{equation*}
	It can be checked that this input distribution satisfies the sum power constraints \eqref{eq:sumpeak} and \eqref{eq:sumaverage}.
	To compute $I(\vect{Z}_1^n; \vect{Y}_1^n)$ for this distribution, we again use Lemma \ref{lemm:PrelovVerdu}. Calculation shows that
	\begin{eqnarray*}
		\frac{1}{n} I(\vect{Z}_1^n; \vect{Y}_1^n ) =  \frac{\rho^2}{2} \sum_{r=1}^{\nr} \left\{ \sum_{t=1}^{\nt} a_t\frac{1}{n} \left( \sum_{1\le i,j \le n}|R_{r,t}(i-j)|^2\right) \right.~ \\ \left.- \left(\sum_{t=1}^{\nt} a_t R_{r,t}(0) \right)^2 \right\} + o(\rho^2).
	\end{eqnarray*}
	Similarly as \eqref{eq:SISO_proof_87}, we have that by \eqref{eq:I_rt_rho},
	\begin{equation*}
		\lim_{n\to\infty} \frac{1}{n} \left( \sum_{1\le i,j \le n}|R_{r,t}(i-j)|^2\right) = \lambda_{r,t} .
	\end{equation*}
	Thus, for every $\vect{a} \in \set{A}(\beta)$, under the input distributions chosen according to $\vect{a}$,
	\begin{eqnarray}
		\lim_{n\to\infty} \lim_{\rho\downarrow 0} \frac{\frac{1}{n} I(\vect{Z}_1^n; \vect{Y}_1^n)}{\rho^2} ~~~~~~~~~~~~~~~~~~~~~~~~~~~~~~~~~~~\nonumber\\
		= \frac{1}{2} \sum_{r=1}^{\nr} \left\{ \sum_{t=1}^{\nt} a_t \lambda_{r,t} - \left(\sum_{t=1}^{\nt} a_t R_{r,t}(0) \right)^2 \right\}.\label{eq:MIMOS_proof_124}
	\end{eqnarray}
	Choosing $\vect{a}$ to be the vector that achieves the maximum in \eqref{eq:MIMOS_proof_122} completes the proof.
\hfill
\IEEEQED

We shall now derive Corollary \ref{coro:MIMOSseparable} from Proposition \ref{prop:MIMOScapacity}. 

\emph{Proof of Corollary \ref{coro:MIMOSseparable}:}
	If the channel is transmit separable, we have $\lambda_{r,t} = \alpha_t^2 \lambda_r$ and $R_{r,t}(0) = \alpha_t R_r(0)$. Equation \eqref{eq:MIMOScapacity} reduces to
	\begin{eqnarray}
		c_\text{S}(\beta) & = & \frac{1}{2} \max_{\vect{a}\in \set{A}(\beta)} \sum_{r=1}^{\nr} \left\{ \lambda_r \sum_{t=1}^{\nt} a_t \alpha_t^2 \right. \nonumber \\ &&~~~~~~~\left. {}- R_r^2(0) \left( \sum_{t=1}^{\nt} a_t \alpha_t \right)^2 \right\}. \label{eq:MIMOSproof3}
	\end{eqnarray}
	Assume without loss of generality that $\alpha_1 \ge \alpha_t$ for all $t\in\{2,\ldots,\nt\}$, i.e., that the first transmit antenna is the strongest. We shall next show that, under this assumption, it is optimal to concentrate all the transmit power on the first antenna. To be more precise, we shall show that for any $\vect{a}$, if $\vect{a}'$ is given by
	\begin{equation}\label{eq:MIMOS_proof_125}
		a'_t = \begin{cases} \frac{\sum_{t=1}^{\nt} a_t \alpha_t}{\alpha_1} & t=1, \\ 0 & \text{otherwise,} \end{cases}
	\end{equation}
	then
	\begin{equation}\label{eq:MIMOS_proof_126}
		\sum_{r=1}^{\nr} \left\{ \lambda_r \sum_{t=1}^{\nt} a_t \alpha_t^2 - R_r^2(0) \left( \sum_{t=1}^{\nt} a_t \alpha_t \right)^2 \right\} \le \sum_{r=1}^{\nr} \left\{ \lambda_r \sum_{t=1}^{\nt} a'_t \alpha_t^2 - R_r^2(0) \left( \sum_{t=1}^{\nt} a'_t \alpha_t \right)^2 \right\}. 
	\end{equation}
	Note that $\vect{a}' \in \set{A}(\beta)$ whenever $\vect{a}\in\set{A}(\beta)$. Inequality \eqref{eq:MIMOS_proof_126} follows because, according to \eqref{eq:MIMOS_proof_125},
	\begin{eqnarray*}
		\sum_{t=1}^{\nt} a_t \alpha_t  & = & \sum_{t=1}^{\nt} a'_t \alpha_t, \\
		\sum_{t=1}^{\nt} a_t \alpha_t^2 & \le & \sum_{t=1}^{\nt} a'_t \alpha_t^2.
	\end{eqnarray*}
	Thus, we conclude that the maximization over $\set{A}(\beta)$ in \eqref{eq:MIMOSproof3} can be reduced to a maximization over the set $\left\{ \vect{a} : 0\le a_1 \le \frac{1}{\beta}; a_t=0, t=2,\ldots,\nt \right\}$. This establishes \eqref{eq:MIMOSseparable}.
\hfill
\IEEEQED

\emph{Proof of Corollary \ref{coro:MIMOSnonephemeral}:}
	In \eqref{eq:MIMOSseparable}, when $\beta=1$ and $\lambda_r \ge 2R_r^2(0)$, the optimal choice of $a$ is $a^*=1$.
\hfill
\IEEEQED


\section{MIMO channels with individual power constraints} \label{sec:MIMOindiv}

In this section we shall prove some capacity bounds for MIMO channels with individual power constraints, and then use these bounds to prove the main results given in Section \ref{sub:MIMOindiv} about such channels.

We shall first give an upper bound on the capacity that is valid for any SNR. To this end, we introduce a few definitions. Let $\set{D}$ be the set of all probability distributions on $\{1,\ldots,\nt\}$:
\begin{equation}
	\set{D} \triangleq \left\{ \vect{d} = \trans{(d_1,\ldots, d_{\nt})}: d_t\ge 0, \  t\in\{1,\ldots,\nt\}; \text{and} \ \sum_{t=1}^{\nt} d_t \le 1\right\}.
\end{equation}
Let $\set{B}$ be the set of all length-$\nt$ binary sequences: 
\begin{equation}
	\set{B} \triangleq \{0,1\}^{\nt}.
\end{equation}
Further, for any $\beta$, let $\set{P}(\beta)$ be a set of probability distributions on $\set{B}$ defined as
\begin{equation}
	\set{P}(\beta) \triangleq \left\{ \vect{p}: \sum_{\vect{b}\in \set{B}}p_{\vect{b}}  b_t  \le \frac{1}{\beta},\ t \in\{1,\ldots,\nt\}\right\}.
\end{equation}
\begin{proposition}\label{prop:MIMOIupper}
	For any $\rho>0$, $\beta\ge 1$ and any $\vect{d} \in \set{D}$,
	\begin{equation}
		C_{\text{I}}(\rho,\beta) \le U_{\text{I}}(\rho,\beta,\vect{d}),\label{eq:MIMOIupper}
	\end{equation}
	where
	\begin{eqnarray}
		U_{\text{I}}(\rho,\beta,\vect{d}) \triangleq \max_{\vect{p} \in \set{P}(\beta)} \sum_{r=1}^{\nr} \left\{ \log \left( 1 + \rho \sum_{\vect{b}\in \set{B}} p_{\vect{b}} \left(\sum_{t=1}^{\nt} b_t R_{r,t}(0) \right) \right) \right. \nonumber \\
		\left. {} + \sum_{\vect{b}\in\set{B}} p_{\vect{b}} \log \left( 1 + \rho \sum_{t=1}^{\nt} b_t \sigma_{r,t}^2 \left(\frac{\rho}{d_t} \right) \right) \right\},\label{eq:MIMOI_def_U}
	\end{eqnarray}
	where $\sigma_{r,t}^2\left(\infty\right)$ is taken to be $0$.
\end{proposition}

The proof of this upper bound is a combination of the proofs for MISO and SIMO channels. We shall give a proof of this bound for MISO channels. The bound for SIMO channels and for general MIMO channels can be proved in exactly the same way as in the sum power constraint case, therefore we omit these two parts.

For MISO channels ($\nr=1$), the above proposition reduces to the following lemma.

\begin{lemma}
	For any $\rho>0$, $\beta\ge 1$ and any $\vect{d} \in \set{D}$,
	\begin{equation}
		C_{\text{MISO-I}} (\rho, \beta) \le U_{\text{MISO-I}}(\rho, \beta, \vect{d}),
	\end{equation}
	where
	\begin{eqnarray}
		U_{\text{MISO-I}}(\rho, \beta, \vect{d}) \triangleq \max_{\vect{p} \in \set{P}(\beta)} \left\{ \log \left( 1 + \rho \sum_{\vect{b}\in \set{B}} p_{\vect{b}} \left(\sum_{t=1}^{\nt} b_t R_{t}(0) \right) \right) \right. \nonumber \\
		\left. {} + \sum_{\vect{b}\in\set{B}} p_{\vect{b}} \log \left( 1 + \rho \sum_{t=1}^{\nt} b_t \sigma_{t}^2 \left(\frac{\rho}{d_t} \right) \right) \right\}.\label{eq:def_U_MISO-I}
	\end{eqnarray}
\end{lemma}
\begin{proof}
	In analogy to the SISO case, to prove \eqref{eq:MIMOIupper}, it suffices to show
	\begin{equation}\label{eq:MIMOI_proof_135}
		I(\vect{Z}_{-\infty}^0, Y_{-\infty}^{-1}; Y_0) \le U_{\text{MISO-I}}(\rho,\beta,\vect{d})
	\end{equation}
	for all input distributions satisfying the individual power constraints \eqref{eq:indivpeak} and \eqref{eq:indivaverage}, and for all $\vect{d}\in\set{D}$. To this end, we follow the proof of Lemma \ref{lemm:SMISO}, but, instead of $W_k'$ as defined in \eqref{eq:SdecomposeW}, we introduce
	\begin{equation*}
		W_k'' = \sum_{t=1}^{\nt} \sqrt{d_t} N_{k}^{(t)}  Z_k^{(t)} + \sqrt{ 1- \sum_{t=1}^{\nt} d_t | Z_k^{(t)} |^2 } N_k' 
	\end{equation*}
	for every $\vect{d}\in\set{D}$ to replace the additive noise $W_k$. Here $\{N_{k}^{(1)}, \ldots, N_{k}^{(\nt)}\}$ and $N_k'$ are defined in the same way as for \eqref{eq:SdecomposeW}, i.e., they are IID random variables of law $\NormalC{0}{1}$ and are independent of the other channel variables. Instead of \eqref{eq:SdecomposeY}, we write $Y_k$ as
	\begin{equation*}
		Y_k = \sum_{t=1}^{\nt} \left(\sqrt{\rho} H_k^{(t)} + \sqrt{d_t} N_k^{(t)} \right) Z_k^{(t)}
		+ \sqrt{ 1- \sum_{t=1}^{\nt} d_t | Z_k^{(t)} |^2 } W_k'.
	\end{equation*}
	Following the steps in the proof of Lemma \ref{lemm:SMISO} we have, instead of \eqref{eq:Suppermutual},
	\begin{eqnarray}
		I(\vect{Z}_{-\infty}^0, Y_{-\infty}^{-1}; Y_0) \le \log \left( 1+ \sum_{t=1}^{\nt} R_t(0) \E{|Z_0^{(t)}|^2} \right)~~~~~\nonumber \\  {}- \E{ \log \left(1+\rho \sum_{t=1}^{\nt} |Z_0^{(t)}|^2\sigma_t^2\left(\frac{\rho}{d_t} \right) \right)}. \label{eq:MIMOI_proof_111}
	\end{eqnarray}
	By the concavity of the $\log$ function, to maximize the RHS of \eqref{eq:MIMOI_proof_111} over distributions on $\vect{Z}_0$, it suffices to consider the case when each input signal has either magnitude zero or one, i.e., it suffices to consider the case when the vector $\trans{\left( |Z_0^{(1)}|^2, \ldots, |Z_0^{(\nt)}|^2 \right)}$ takes value in $\set{B}$. Let $\vect{p}$ be the probability distribution of $\trans{\left( |Z_0^{(1)}|^2, \ldots,|Z_0^{(\nt)}|^2 \right)}\in\set{B}$. Note that, according to the individual average-power constraint \eqref{eq:indivaverage}, $\vect{p}$ must satisfy $\vect{p} \in \set{P}(\beta)$. We thus obtain that maximizing the RHS of \eqref{eq:MIMOI_proof_111} yields $U_{\text{MISO-I}}(\rho,\beta,\vect{d})$ as defined in \eqref{eq:def_U_MISO-I}. This establishes \eqref{eq:MIMOI_proof_135}.
\end{proof}

The next corollary, which follows from Proposition \ref{prop:MIMOIupper}, gives an upper bound on the asymptotic capacity of MIMO channels with individual power constraints.

\begin{corollary}\label{coro:MIMOIupper}
	For any $\beta\ge 1$ and $\vect{d}\in\set{D}$ satisfying $d_t>0$, $t \in \{ 1,\ldots, \nt\}$,
	\begin{eqnarray}
		\limsup_{\rho\downarrow 0} \frac{C_{\text{I}}(\rho,\beta)}{\rho^2} \le \max_{\vect{p} \in \set{P}(\beta) } \frac{1}{2} \sum_{r=1}^{\nr}~~~~~~~~~~~~~~~~~~~~~~~~~~~\nonumber \\
		 \left\{ \sum_{\vect{b} \in \set{B}} p_{\vect{b}} \left( \sum_{t=1}^{\nt} b_t \cdot \frac{\lambda_{r,t} - R_{r,t}^2(0)}{d_t} + \left(\sum_{t=1}^{\nt} b_t R_{r,t}(0) \right)^2 \right) \right. \nonumber \\
		\left. {} - \left(\sum_{\vect{b}\in \set{B} } p_{\vect{b}} \left(\sum_{t=1}^{\nt} b_t R_{r,t}(0) \right) \right)^2 \right\}.~~~~~~~~~~ \label{eq:MIMOIupper2}
	\end{eqnarray}
\end{corollary}

\begin{proof}
Inequality \eqref{eq:MIMOIupper2} follows from \eqref{eq:MIMOIupper} and \eqref{eq:MIMOI_def_U} by application of the second order Taylor Expansion of the function $x\mapsto\log(1+x)$ and the first order Taylor Expansion of $\sigma_{r,t}^2 (\cdot)$. The latter is given by
\begin{equation}\label{eq:Taylorsigma}
	\sigma_{r,t}^2(\rho) = R_{r,t}(0) - \frac{\lambda_{r,t} - R_{r,t}^2(0)}{2} \rho + o(\rho),
\end{equation}
which can be obtained from \eqref{eq:sigma_lambda}.
\end{proof}

\emph{Proof of Proposition \ref{prop:MIMOIcapacity}:}
	We first consider the upper bound \eqref{eq:MIMOIupper2} for transmit separable channels. For such channels, we have $R_{r,t}(0) = \alpha_t R_r(0)$ and $\lambda_{r,t} = \alpha_t^2 \lambda_r$. Choosing $\vect{d}$ to be
	\begin{equation*}
		d_t = \frac{\alpha_t}{\sum_{s=1}^{\nt} \alpha_s},\quad t \in \{1,\ldots,\nt\},
	\end{equation*}
	and denoting
	\begin{equation}\label{eq:MIMOI_proof_140a}
		a(\vect{p}) \triangleq \frac{\sum_{\vect{b} \in \set{B}} p_{\vect{b}} \sum_{t=1}^{\nt} b_t \alpha_t}{\sum_{t=1}^{\nt} \alpha_t},
	\end{equation}
	\begin{eqnarray}
		\limsup_{\rho\downarrow 0} \frac{C_{\text{I}}(\rho,\beta)}{\rho^2} \le \max_{\vect{p} \in \set{P}(\beta)} \frac{1}{2} \left(\sum_{t=1}^{\nt} \alpha_t \right)^2 \sum_{r=1}^{\nr}~~~~~~~~~~~ \nonumber\\
		\Bigg\{ \left(\lambda_r - R_r^2(0) \right) a(\vect{p}) - R_r^2(0) a^2(\vect{p})~~~~~~~~  \nonumber\\
		 {} + R_r^2(0) \left(\sum_{\vect{b} \in \set{B}} p_{\vect{b}} \left(\frac{\sum_{t=1}^{\nt} b_t \alpha_t}{\sum_{t=1}^{\nt} \alpha_t}\right)^2\right)\Bigg\}.\label{eq:MIMOI_proof_139}
	\end{eqnarray}
	Note that since
	\begin{equation}\label{eq:MIMOI_proof_142a}
		\frac{\sum_{t=1}^{\nt} b_t \alpha_t}{\sum_{t=1}^{\nt} \alpha_t} \le \frac{\sum_{t=1}^{\nt} \alpha_t}{\sum_{t=1}^{\nt} \alpha_t} = 1,
	\end{equation}
	the square of the LHS of \eqref{eq:MIMOI_proof_142a} is less than or equal to itself. Thus from \eqref{eq:MIMOI_proof_140a} and \eqref{eq:MIMOI_proof_142a} we have
	\begin{equation}\label{eq:MIMOI_proof_143a}
		\sum_{\vect{b}\in\set{B}} p_{\vect{b}} \left( \frac{\sum_{t=1}^{\nt} b_t\alpha_t}{\sum_{t=1}^{\nt} \alpha_t} \right)^2 \le a(\vect{p}).
	\end{equation}
	From \eqref{eq:MIMOI_proof_139} and \eqref{eq:MIMOI_proof_143a} we obtain
	\begin{equation}\label{eq:MIMOI_proof_140}
		\limsup_{\rho\downarrow 0} \frac{C_{\text{I}}(\rho,\beta)}{\rho^2} \le \frac{1}{2} \left( \sum_{t=1}^{\nt} \alpha_t \right)^2 \max_{\vect{p} \in \set{P}(\beta)} \sum_{r=1}^{\nr} \left\{a(\vect{p}) \lambda_r - a^2(\vect{p}) R_r^2(0) \right\}.
	\end{equation}
	Noting that the RHS of \eqref{eq:MIMOI_proof_140} depends on $\vect{p}$ only through $a(\vect{p})$, we replace $a(\vect{p})$ by $a$. Then \eqref{eq:MIMOI_proof_140} reduces to
	\begin{equation}\label{eq:MIMOI_proof_141}
		\limsup_{\rho\downarrow 0} \frac{C_{\text{I}}(\rho, \beta)}{\rho^2} \le \frac{1}{2} \left(\sum_{t=1}^{\nt} \alpha_t\right)^2 \max_{0\le a \le \frac{1}{\beta}} \sum_{r=1}^{\nr} \left\{ a \lambda_r - a^2 R_r^2(0) \right\}.
	\end{equation}
	The RHS of \eqref{eq:MIMOI_proof_141} is the same as that of \eqref{eq:MIMOIcapacity}. 
	
	To derive a lower bound on the asymptotic capacity, we propose an input distribution on $\vect{Z}_1^n$ for every $n\in\Naturals$. Such distributions are given by
	\begin{equation*}
		Z_k^{(1)} = \cdots = Z_k^{(\nt)} = U \cdot \Phi_k,\quad k \in \{1,\ldots,n\},
	\end{equation*}
	where $U$ and $\Phi_k$ are chosen in the same way as for SISO channels, as described in Section \ref{sec:SISO}. We apply Lemma \ref{lemm:PrelovVerdu} to obtain	\begin{equation}\label{eq:MIMOI_proof_142}
		\lim_{n\to \infty} \lim_{\rho\downarrow 0} \frac{\frac{1}{n} I(\vect{Z}_1^n; \vect{Y}_1^n)}{\rho^2} = \frac{1}{2} \left(\sum_{t=1}^{\nt} \alpha_t\right)^2 \max_{0\le a \le \frac{1}{\beta}} \sum_{r=1}^{\nr} \left\{ a \lambda_r - a^2 R_r^2(0) \right\}.
	\end{equation}
	By Lemma \ref{lem:lower}, this forms a lower bound on
	\begin{equation*}
		\liminf_{\rho\downarrow 0} \frac{C_{\text{I}}(\rho,\beta)}{\rho^2}.
	\end{equation*}
	Combining \eqref{eq:MIMOI_proof_141} and \eqref{eq:MIMOI_proof_142} establishes \eqref{eq:MIMOIcapacity}.
\hfill
\IEEEQED

\emph{Proof of Corollary \ref{coro:MIMOInonephemeral}:}
Note that when $\beta=1$ and the channel is nonephemeral, i.e., $\lambda_r \ge 2 R_r^2(0)$ for all $r \in \{1,\ldots,\nr\}$, the choice of $a$ that maximizes the RHS of \eqref{eq:MIMOIcapacity} is $a=1$. Thus, in this case, \eqref{eq:MIMOIcapacity} reduces to \eqref{eq:MIMOInonephemeral}.
\hfill
\IEEEQED

For channels that are nonephemeral, with no average-power constraint ($\beta=1$), but that are not necessarily transmit separable, we have the following upper and lower bounds that in general do not coincide.

\begin{corollary}
	If the MIMO channel is nonephemeral and if no average-power constraint is imposed ($\beta=1$), we have an upper bound on the capacity asymptote given by
	\begin{equation}\label{eq:MIMOI_loose_upper}
		\limsup_{\rho\downarrow 0} \frac{C_{\text{I}}(\rho,\beta)}{\rho^2} \le \nt \sum_{r=1}^{\nr} \sum_{t=1}^{\nt} \frac{\lambda_{r,t} - R_{r,t}^2(0)}{2}
	\end{equation}
	and a lower bound given by
	\begin{equation}\label{eq:MIMOI_loose_lower}
		\liminf_{\rho\downarrow 0} \frac{C_{\text{I}}(\rho,\beta)}{\rho^2} \ge \sum_{r=1}^{\nr} \sum_{\nu=1}^{\infty} \left| \sum_{t=1}^{\nt} R_{r,t}(\nu)\right|^2.
	\end{equation}
\end{corollary}

\begin{proof}
The upper bound \eqref{eq:MIMOI_loose_upper} is obtained by choosing $\vect{d} = \frac{1}{\nt} \trans{(1,\ldots,1)}$ in \eqref{eq:MIMOIcapacity}. The lower bound \eqref{eq:MIMOI_loose_lower} is obtained by using the input distributions given in the proof of Proposition \ref{prop:MIMOIcapacity}, with $U=1$ with probability one.
\end{proof}

\section{SISO channels with delay spread} \label{sec:delay}

In this section we shall prove Proposition \ref{prop:delaycapacity} and Corollary \ref{coro:delaynonephemeral}. 

\emph{Proof of Proposition \ref{prop:delaycapacity}:}
	As shown in Section \ref{sub:delay}, the capacity of the SISO channel with delay spread \eqref{eq:delaymodel} is upper bounded by the capacity of the MISO channel \eqref{eq:delayMISO} with the same individual peak- and average-power constraints. The latter is obtained by choosing $\nr=1$ in \eqref{eq:MIMOIcapacity}, which yields the same value as the RHS of \eqref{eq:delaycapacity}. Thus, to prove \eqref{eq:delaycapacity}, it only remains to find a lower bound on the asymptotic capacity that coincides with its RHS.
	
	For every $n\in\Naturals$, consider the following distribution on the input signals $Z_1^n$. Let
	\begin{equation*}
		Z_k = U \cdot \Phi_k, \quad k \in \{1,\ldots,n\},
	\end{equation*}
	where $U$ is equal to $1$ with probability $a$ and is equal to $0$ with probability $(1-a)$; $\{\Phi_k\}$ is chosen such that $\Phi_k = \exp(i \cdot k \Theta)$, where $i=\sqrt{-1}$; $\Theta$ is uniformly distributed over the set $\left\{\frac{j\cdot 2 \pi}{m}: j \in \{ 0, \ldots, m-1\} \right\}$ for some $m>1$. The asymptotic value of $I(Z_1^n; Y_1^n)$ for this input distribution is calculated using Lemma \ref{lemm:PrelovVerdu} to yield
	\begin{equation*}
		\lim_{n\to\infty}\lim_{\rho\downarrow 0} \frac{\frac{1}{n}I(Z_1^n; Y_1^n)}{\rho^2}=\frac{1}{2} \left(\sum_{t=0}^{T-1} \alpha_t\right)^2 \max_{0\le a\le \frac{1}{\beta}} \left\{ a \lambda - a^2 R^2(0) \right\}.
	\end{equation*}
	By Lemma \ref{lem:lower}, this gives us the desired lower bound on the asymptotic capacity.
\hfill
\IEEEQED

\emph{Proof of Corollary \ref{coro:delaynonephemeral}:}
When the channel is nonephemeral ($\lambda \ge 2R^2(0)$) and $\beta=1$, the choice of $a$ that maximizes the RHS of \eqref{eq:delaycapacity} is $a=1$.
\hfill
\IEEEQED


\appendices

\section{A Lower Bound on the Capacity}
In this section we present a general lower bound on the capacity of the SISO fading channel considered in this paper. This lemma can be extended to MIMO channels with sum power constraints or individual power constraints, and to SISO channels with delay spread. The proofs for these three cases are exactly the same as for SISO channels.

\begin{lemma}[Lower Bound on Capacity] \label{lem:lower}
	For any $n\in\Naturals$ and any distribution on $Z_1^n$ satisfying the peak- and average-power constraints \eqref{eq:SISOpeak} and \eqref{eq:SISOaverage},
	\begin{equation}\label{eq:lemmlower}
		C(\rho,\beta) \ge \frac{1}{n}I(Z_1^n; Y_1^n).
	\end{equation}
\end{lemma}
\begin{proof}
We extend the distribution on $Z_1^n$ to a distribution on $\{Z_k\}$ in such way that the length-$n$ blocks of input symbols $\{Z_{kn+1}^{(k+1)n}\}_{k=0}^\infty$ are IID according to the law of $Z_1^n$. Clearly, if the given distribution on $Z_1^n$ satisfies Constraints \eqref{eq:SISOpeak} and \eqref{eq:SISOaverage}, then so does the induced distribution on $\{Z_k\}$. We next show that under this distribution on $\{Z_k\}$,
\begin{equation}\label{eq:lower_proof_1}
	\lim_{N\to \infty} \frac{1}{N}I(Z_1^N; Y_1^N) \ge \frac{1}{n} I(Z_1^n; Y_1^n),
\end{equation}
from which the Lemma \ref{lem:lower} follows. To prove \eqref{eq:lower_proof_1}, we let $m\triangleq \lfloor N/n \rfloor$ and write
\begin{eqnarray}
	\frac{1}{N} I(Z_1^N; Y_1^N) & \ge & \frac{1}{N} I(Z_1^{nm}; Y_1^N) \nonumber\\
	& = & \frac{1}{N}\sum_{k=0}^{m-1} I\left( Z_{kn+1}^{(k+1)m}; Y_1^N | Z_1^{kn} \right) \nonumber\\
	& = & \frac{1}{N} \sum_{k=0}^{m-1} I \left( Z_{kn+1}^{(k+1)m}; Y_1^N, Z_1^{kn} \right) \nonumber\\
	& \ge & \frac{1}{N} \sum_{k=0}^{m-1} I \left( Z_{kn+1}^{(k+1)n}; Y_{kn+1}^{(k+1)n} \right) \nonumber\\
	& = & \frac{m}{N} I(Z_1^n; Y_1^n),\label{eq:lower_proof_2}
\end{eqnarray}
where the first inequality follows by omitting terms in the mutual information; the next equality by the chain rule; the next equality because the input symbols in different blocks are mutually independent; the next inequality again by omitting terms in the mutual information; and the last equality because every block of input symbols has the same distribution as the first block ($k=0$). Inequality \eqref{eq:lower_proof_1} follows from \eqref{eq:lower_proof_2} because
\begin{equation*}
	\lim_{N\to\infty} \frac{\lfloor N/n \rfloor}{N} = \frac{1}{n}.
\end{equation*}
\end{proof}


\section{Second-order asymptotics of mutual information}
In this section we restate a special case of  \cite[Corollary 1]{PrelovVerdu04}. Consider  the following channel
\begin{equation}
	\vect{Y} = \sqrt{\rho}\vmat{H} \vect{Z} + \vect{W}
\end{equation}
where $\vect{Y}$, $\vect{Z}$ and $\vect{W}$ are random $n$-vectors and $\vmat{H}$ is an $n\times n$ random matrix. The entries of $\vmat{H}$ can be correlated with each other, and are assumed to be of mean zero and jointly PCN. The coordinates of the additive noise vector $\vect{W}$ are IID random variables of law $\NormalC{0}{1}$.
\begin{lemma}\label{lemm:PrelovVerdu}
	If there exist $\delta_0>0$ and $\nu>0$ such that
	\begin{equation}\label{eq:PrelovVerducondition}
		\Pr\left[\| \vect{Z} \|_2 > \delta\right] \le \exp\{-\delta^\nu\}
	\end{equation}
	for all $\delta>\delta_0$, then
	\begin{eqnarray*}
		I(\vect{Z}; \vect{Y}) = \frac{\rho^2}{2} \text{trace}\left\{ \E{ \left( \E{ \vmat{H}\vect{Z}\vect{Z}^\dagger\vmat{H}^\dagger|\vect{Z}}\right)^2}~~~~~~~~~~~~~~\right. \\ \left. - \left(\E{\vmat{H}\E{\vect{Z}\vect{Z}^\dagger}\vmat{H}^\dagger}\right)^2\right\}+o(\rho^2). 
	\end{eqnarray*}
\end{lemma}

Note that since in this paper we are considering channels with hard peak-power constraints (\eqref{eq:SISOpeak} or \eqref{eq:sumpeak} or \eqref{eq:indivpeak}), Condition \eqref{eq:PrelovVerducondition} is always satisfied.


\section{Proof of Lemma \ref{lemm:predictZY}} \label{app:predictZY}

In this section we shall prove Lemma \ref{lemm:predictZY}. 
Define $Q$ by  $Q \triangleq \sum_{\nu=-\infty}^{\infty} |R(\nu)|$. According to the assumptions of Lemma \ref{lemm:predictZY}, we have that
$Q$ is finite, that $R(0)=1,$  and that the past inputs $ z_{-\infty}^{-1} $ satisfy the peak
power constraint \eqref{eq:SISOpeak}.
Let $\vect{K}$ denote the infinite matrix,  with rows and columns indexed by the negative integers,
with row-$\mu$ column-$\nu$ entry $R(\mu-\nu)$ for negative integers $\mu$ and $\nu$.  
Let $\vect{D}$ denote the infinite diagonal matrix with row-$\mu$ column-$\mu$  entry $z_{\mu}$, for negative integers $\mu$.
Let $\vect{v}$ be the infinite column vector with $\mu$-th entry $R(\mu)$ for negative integers $\mu$.
For $ z_{-\infty}^{-1} $ fixed,  $\vect{I} + \rho\vect{D}\vect{K}\vect{D}^\dagger$ is the covariance
matrix of the observation $Y_{-\infty}^{-1}$, and
$\vect{v}^\dagger \vect{D}^\dagger$ is the covariance between for the variable to be estimated, $H_0$,
and the observation  $Y_{-\infty}^{-1}.$
Although $\vect{K}$ is an infinite matrix, its powers $\vect{K}^j$ are well defined
in terms of absolutely convergent sums.   Indeed, it is easy to show by induction on
$j$ that $ \max_{\mu, \nu} |  (\vect{K}^j)_{\mu,\nu}| \leq Q^{j-1}$ for any $j\geq 1$.
In view of this fact,  for sufficiently small $\rho$,
$(\vect{I}+\rho \vect{D}  \vect{K} \vect{D}^\dagger )^{-1}$  is well defined by
an absolutely convergent series:
\begin{equation} \label{eq:inverse_series}
(\vect{I}+\rho \vect{D} \vect{K} \vect{D}^\dagger )^{-1} =  \vect{I} +  \sum_{j=1}^\infty (-\rho \vect{D} \vect{K} \vect{D}^\dagger  )^j.
\end{equation}
The orthogonality principle can be used to check that the optimal estimator
can be represented by
\begin{equation} \nonumber  
\widehat{H}_0=\vect{v}^\dagger \vect{D}^\dagger  (\vect{I}+\rho \vect{D} \vect{K} \vect{D}^\dagger )^{-1}   Y_{-\infty}^{-1},
\end{equation}
with the minimum mean square error given by
\begin{equation} \label{eq:prediction_error}
\varsigma^2(\rho, z_{-\infty}^{-1})  = 1 - \rho \vect{v}^\dagger \vect{D}^\dagger  (\vect{I}+\rho \vect{D}\vect{K} \vect{D}^\dagger)^{-1} \vect{D}  \vect{v}.
\end{equation}
Substituting \eqref{eq:inverse_series} into \eqref{eq:prediction_error} yields:
$$
\varsigma^2(\rho, z_{-\infty}^{-1})  = 1 - \rho \sum_{\nu=-\infty}^{-1} |R(\nu)|^2 |z_\nu|^2  + \triangle
$$
where
$$
\triangle =   - \rho \vect{v}^\dagger \vect{D}^\dagger \left(  \sum_{j=1}^\infty (-\rho \vect{D}^\dagger \vect{K} \vect{D} )^j \right) \vect{D}\vect{v}.
$$
Let $|\vect{K}|$ be the matrix obtained by replacing each entry of $\vect{K}$ by its magnitude,
and define $|\vect{v}|$ and $|\vect{D}|$ similarly.    Note that $|\vect{D}| \leq \vect{I}$, and the sum of the entries
of $\vect{v}$ is less than or equal to $Q$.   Therefore, for $0 \leq \rho < 1/Q$,
\begin{eqnarray*}
| \triangle |  &  \leq  &  \rho |\vect{v}|^T |\vect{D}| \left(  \sum_{j=1}^\infty (\rho |\vect{D}| |\vect{K}| |\vect{D}| )^j \right) |\vect{D}| |\vect{v}|  \\
& \leq &       \sum_{j=1}^\infty  \rho   |\vect{v}|^T (\rho  |\vect{K}|  )^j   |\vect{v}|   \\
& \leq &      \sum_{j=1}^\infty     \rho^{j+1}  Q^{j+1}  = \frac{\rho^2 Q^2}{1-\rho Q} = o(\rho).
\end{eqnarray*}
Lemma \ref{lemm:predictZY} is proved.


\bibliographystyle{ieeetr}

\begin{IEEEbiographynophoto}{Vignesh Sethuraman}
 received a B.Tech. from Indian Institute of Technology, Madras, in 2001, an M.S. and a Ph.D. from University of Illinois at Urbana-Champaign in 2003 and 2006. He is currently with Qualcomm. His research interests include wireless communication and information theory. 
\end{IEEEbiographynophoto}
\begin{IEEEbiographynophoto}{Ligong Wang} 
 (Student Member '08) received the degree of Bachelor of Engineering from Tsinghua University in 2004 and the degree of Master of Science from ETH Zurich in 2006. He is currently a doctoral student at the ETH Zurich. His research interests are in classical and quantum information theory. In particular, he is interested in fading channels, the Poisson channel, and finite block-length problems.
\end{IEEEbiographynophoto}
\begin{IEEEbiographynophoto}{Bruce Hajek} 
(M'79-SM'84-F'89) received a B.S. in Mathematics and an M.S. in Electrical Engineering from the University of Illinois at Urbana-Champaign in 1976 and 1977, and a Ph. D. in Electrical Engineering from the University of California at Berkeley in 1979. He is a Professor in the Department of Electrical and Computer Engineering and in the Coordinated Science Laboratory at the University of Illinois at Urbana-Champaign, where he has been since 1979. His research interests include communication and computer networks, stochastic systems, combinatorial and nonlinear optimization, and information theory. He served as Associate Editor for Communication Networks and Computer Networks for the IEEE Transactions on Information Theory, as Editor-in-Chief  of the same Transactions, and as President of the IEEE Information Theory Society. He is a member of the US National Academy of Engineering and he was a winner of the 1973 USA Mathematical Olympiad. He received the Eckman Award of the American Automatic  Control  Council, an NSF Presidential Young Investigator Award, an Outstanding Paper Award from the IEEE Control Systems Society,  and  the IEEE Kobayashi Computer Communications Award.
\end{IEEEbiographynophoto}
\begin{IEEEbiographynophoto}{Amos Lapidoth}
 (S'89--M'95--SM'00--F'04) received the B.A.\ degree in mathematics
 (\emph{summa cum laude}, 1986), the B.Sc.\ degree in electrical
 engineering (\emph{summa cum laude}, 1986), and the M.Sc.\ degree in
 electrical engineering (1990), all from the Technion--Israel
 Institute of Technology, Haifa. He received the Ph.D.\ degree in
 electrical engineering from Stanford University, Stanford, CA, in
 1995.

 During 1995--1999, he was an Assistant and Associate Professor in
 the Department of Electrical Engineering and Computer Science at the
 Massachusetts Institute of Technology, Cambridge, and was the KDD
 Career Development Associate Professor in Communications and
 Technology. He is now Professor of Information Theory at the Signal
 and Information Processing Laboratory, ETH Zurich, Switzerland. His
 research interests are in digital communications and information
 theory.
\end{IEEEbiographynophoto}

\end{document}